# Multiple testing with the structure adaptive Benjamini-Hochberg algorithm

Ang Li and Rina Foygel Barber

**Abstract**

In multiple testing problems, where a large number of hypotheses are tested simultaneously, false discovery rate (FDR) control can be achieved with the well-known Benjamini-Hochberg procedure, which adapts to the amount of signal present in the data, under certain distributional assumptions. Many modifications of this procedure have been proposed to improve power in scenarios where the hypotheses are organized into groups or into a hierarchy, as well as other structured settings. Here we introduce SABHA, the "structure-adaptive Benjamini-Hochberg algorithm", as a generalization of these adaptive testing methods. SABHA incorporates prior information about any pre-determined type of structure in the pattern of locations of the signals and nulls within the list of hypotheses, to reweight the p-values in a data-adaptive way. This raises the power by making more discoveries in regions where signals appear to be more common. Our main theoretical result proves that SABHA controls FDR at a level that is at most slightly higher than the target FDR level, as long as the adaptive weights are constrained sufficiently so as not to overfit too much to the data—interestingly, the excess FDR can be related to the Rademacher complexity or Gaussian width of the class from which we choose our data-adaptive weights. We apply this general framework to various structured settings, including ordered, grouped, and low total variation structures, and get the bounds on FDR for each specific setting. We also examine the empirical performance of SABHA on fMRI activity data and on gene/drug response data, as well as on simulated data.

## 1 Introduction

In modern scientific fields with high-dimensional data, the problem of multiple testing arises whenever we search over a large number of questions or hypotheses with the hope of discovering signals while maintaining error control. Treating the many questions separately may lead to a large list of discoveries which are mostly spurious; instead, error measures such as the family-wise error rate (FWER) and the false discovery rate (FDR) are popular tools for producing a more reliable and replicable list of discoveries based on the available data.

To formalize the problem, consider a list of $n$ hypotheses $H_1, \ldots, H_n$, accompanied by p-values $P_1, \ldots, P_n \in [0, 1]$ which have been computed based on observed data. Classical approaches to the multiple testing problem generally treat these $n$ questions exchangeably; our only information about hypothesis $i$ comes from its p-value $P_i$. In many settings, however, we may have prior information or beliefs about the pattern of signals and nulls among these $n$ questions. For example, we may believe that certain hypotheses are more likely to contain signals than others (e.g. due to data from prior experiments); or that the true signals are likely to be clustered (e.g. if each hypothesis corresponds to a test performed at some spatial location, and the true signals are likely to appear in spatial clusters); or the hypotheses may come with some natural grouping, with true signals tending to co-occur in the same group (e.g. if each hypothesis corresponds to a gene, then known gene pathways may form such a grouping).

In this work, we give a general framework for incorporating this type of information when performing multiple testing. We introduce the *structure-adaptive Benjamini-Hochberg procedure*, a procedure which places data-adaptive weights on the p-values in order to adapt to the apparent patterns of signals and nulls in the data. This procedure offers increased power to detect signals by lowering the threshold for making discoveries in regions where the data suggests that signals are highly likely to occur. When run with a target FDR level $\alpha$, the method offers a finite-sample guarantee of FDR control at a level that is only slightly higher than $\alpha$ as long as we restrict the extent to which the weights can adapt to the data, thus avoiding overfitting.

**Outline**   In Section 2 we give background on the multiple testing problem, including some existing methods to accommodate known patterns in the signals and nulls, and then present the SABHA method. Section 3 contains our main theoretical results on FDR control for this method, for independent or dependent p-values. In Section 4 we give the details for applying our method in several different settings. Our main theoretical results are proved in Section 5. In Section 6, we then present empirical results on simulated data and on several real data sets, with all code to reproduce all experiments available online.[1] We conclude with a discussion of our findings and directions for future work in Section 7. Some proofs and calculations are deferred to the Appendix.

## 2  Method

We first introduce the multiple testing problem and various structured methods for false discovery rate (FDR) control in Section 2.1, then introduce our method in Section 2.2.

### 2.1  Background: multiple testing and FDR control

**False discovery rates and the Benjamini-Hochberg (BH) procedure**   Given a multiple testing problem consisting of p-values $P_1, \ldots, P_n$ corresponding to $n$ hypotheses, Benjamini and Hochberg [6] propose a procedure for setting an adaptive threshold for these p-values, which seeks to control the false discovery rate (FDR) rather than more conservative error measures. The FDR is defined as the expected false discovery proportion,

$$\text{FDR} = \mathbb{E}\left[\text{FDP}\right] \text{ where } \text{FDP} = \frac{\sum_i \mathbb{1}\left\{P_i \text{ rejected and } i \in \mathcal{H}_0\right\}}{1 \vee \sum_i \mathbb{1}\left\{P_i \text{ rejected}\right\}}.$$

Here $\mathcal{H}_0 \subseteq [n] := \{1, \ldots, n\}$ is the (unknown) set of null hypotheses, i.e. the p-values $P_i$ for $i \in \mathcal{H}_0$ correspond to testing hypotheses where no signal is present (often, we have $P_i$ uniformly distributed for $i \in \mathcal{H}_0$, but in some settings, e.g. using discrete data, this may not be exactly the case.)

The BH procedure proceeds by calculating

$$\widehat{k} = \max\left\{k \geq 1 : P_i \leq \alpha \cdot \frac{k}{n} \text{ for at least } k \text{ many p-values } P_i\right\},$$

or setting $\widehat{k} = 0$ if this set is empty, then rejecting any p-value $P_i$ which satisfies $P_i \leq \alpha \cdot \frac{\widehat{k}}{n}$, for a total of $\widehat{k}$ many rejections. We can think of this as setting an adaptive rejection threshold, $\alpha \cdot \frac{\widehat{k}}{n}$, which lies between the naive threshold $\alpha$ and the Bonferroni adjusted threshold $\alpha/n$.

In the setting where the null p-values are uniformly distributed, are mutually independent, and are independent of the non-null p-values, Benjamini and Hochberg [6] prove that the BH procedure controls the FDR at the level

$$\text{FDR} = \alpha \cdot \frac{|\mathcal{H}_0|}{n} \leq \alpha.$$

---

[1] Code available at http://www.stat.uchicago.edu/~rina/sabha.html.

In fact, under some types of dependence, the FDR is again bounded by $\alpha$; for example, if the p-values are positively dependent (Benjamini and Yekutieli [7]'s PRDS condition), or in an asymptotic regime where the empirical distribution of the nulls resembles the uniform distribution [28].

**Modifying the BH procedure by estimating the proportion of signals**  Looking at the above bound $\alpha \cdot \frac{|\mathcal{H}_0|}{n}$ on the FDR, we see that in situations where the number of signals is a substantial proportion of the total number of hypotheses, the Benjamini-Hochberg procedure will be overly conservative. To adapt to the proportion of signals, Storey [27] proposes a two-stage method where first, the relative proportion of high and low p-values is used to estimate the proportion of nulls:

$$\widehat{\pi}_0 = \min\left\{1, \frac{\sum_i \mathbb{1}\{P_i > \tau\}}{n(1-\tau)}\right\}. \tag{1}$$

(To understand this intuitively, if $\pi_0 = \frac{|\mathcal{H}_0|}{n}$ is the true proportion of nulls, and if the signals $i \notin \mathcal{H}_0$ are all strong with $\Pr(P_i > \tau) \approx 0$, then we should have approximately $\pi_0 n \cdot (1-\tau)$ many p-values which are greater than $\tau$.) Second, this estimated proportion is used to adjust the threshold in the Benjamini-Hochberg procedure: by running BH with $\frac{\alpha}{\widehat{\pi}_0}$, we now expect the FDR to be

$$\text{FDR} \approx \frac{\alpha}{\widehat{\pi}_0} \cdot \frac{|\mathcal{H}_0|}{n} \lessapprox \alpha,$$

where the first step holds by the FDR level of the BH procedure, while the second holds when $\widehat{\pi}_0$ is a good (over)estimate of $\pi_0 = \frac{|\mathcal{H}_0|}{n}$. Storey et al. [28, Theorem 3] proves that this procedure, with a slightly more conservative definition of $\widehat{\pi}_0$ and rejecting only those p-values no larger than $\tau$, controls the FDR at the desired level $\alpha$.

In some settings, we may believe that there is asymmetry in the true signals in terms of positive versus negative effects—if the p-values are computed from real-valued statistics whose null distribution is symmetric (e.g. a z-score), and the nonnulls are more likely to have positive means than negative (or vice versa), then taking this into account in the adaptive procedure will increase power. Zhao and Fung [30] allow for this asymmetry by giving distinct weights to p-values associated to positive versus negative statistics, specifically, by estimating $\widehat{\pi}_0^+$ and $\widehat{\pi}_0^-$, the proportion of nulls among hypotheses with positive or negative z-scores, respectively. More generally, Sun and Cai [29] study oracle decision rules for discovering signals, and find that using z-scores (i.e. a signed statistic) can outperform a p-value based procedure by making use of the sign information.

**FDR control with structured signals and nulls**  Here we briefly describe several existing methods that adapt to particular types of structure in the signals and nulls; more detailed comparisons with the methods most relevant to our work will be given later on.

First, Hu et al. [17] consider a setting where the hypotheses are partitioned into disjoint groups $\mathcal{G}_1, \ldots, \mathcal{G}_d \subseteq [n]$. The groups may differ widely in terms of the proportion of nulls within the group, $\pi_0^{(k)} = \frac{|\mathcal{H}_0 \cap \mathcal{G}_k|}{|\mathcal{G}_k|}$. In their "Group Benjamini-Hochberg" procedure, after estimating these proportions with some $\widehat{\pi}_0^{(k)}$'s, the p-values are then reweighted to take these estimates into account before applying the BH procedure; this allows for higher power to make discoveries within those groups where signals are prevalent. This can be viewed as an extension of the work of Storey [27], which estimates the overall proportion of nulls $\pi_0$ without separating into groups. Hu et al. [17]'s main theoretical results show, firstly, that if the true $\pi_0^{(k)}$'s are known exactly, then FDR is controlled at the desired level $\alpha$, and second, that if the $\pi_0^{(k)}$'s can instead be estimated consistently, then asymptotic FDR control is achieved. It is important to note that in this setting, the goal is not to treat each group as a single unit that is rejected or not—instead, the rejections are carried out at the level of the individual hypotheses, but the predetermined group structure helps to increase power by identifying groups where there are high proportions of non-nulls.

In comparison, cluster-wise FDR methods search for data-determined clusters of signals within the data, often treating a cluster (i.e. a data-determined continguous block of rejection) as a single "discovery" for the purpose of defining FDR. This type of multiple testing with spatial correlations or clustering has been extensively studied and arises in many applications, for instance, genomics, geophysical sciences, and astronomy. Chouldechova [11], Schwartzman et al. [24]. and Cheng and Schwartzman [10] develop a cluster-wise FDR, which treats discoveries at the cluster level rather than counting individual points; in Chouldechova [11] this is done by modeling the detected cluster sizes for null versus non-null clusters in order to adjust the usual point-wise FDR bounds and obtain cluster-wise FDR guarantees, while Schwartzman et al. [24] proceed by testing only at the local maxima of the sequence (for a one-dimensional spatial setting), and Cheng and Schwartzman [10] develop this idea further by testing extreme values of the *derivative* of the sequence, rather than the values of the sequence itself. The problem of detecting data-determined clusters of non-nulls is studied also by Siegmund et al. [25] in the context of copy number variation in genetics.

Next, in some settings we may have reason to reject the hypotheses in an ordered way, that is, to search for a cutoff point in our list so that $P_1, \ldots, P_k$ are labeled as likely signals and $P_{k+1}, \ldots, P_n$ as likely nulls (i.e. $P_1, \ldots, P_k$ are our "discoveries"). This may be desirable either when prior information suggests which hypotheses are most likely to contain signals—in which case we can place these at the beginning of our list before gathering data—or when the structure of the problem demands it, for instance, in a model selection algorithm where $P_k$ corresponds to the variable selected at time $k$, it might be counterintuitive to conclude that $P_{k+1}$ contains a true signal while $P_k$ does not. G'Sell et al. [16], Barber and Candès [1], and and Li and Barber [21] each propose various methods for this ordered testing problem with accompanying FDR control guarantees; a related method, which does not require a contiguous block of rejections, is the Adaptive SeqStep method of Lei and Fithian [20].

Another structure commonly found in the literature is when the p-values are grouped into a hierarchy, with methods in this setting developed by Benjamini and Bogomolov [5] and applied to fMRI data by Schildknecht et al. [23].

Each of the above settings can be viewed as a scenario where the multiple testing procedure adapts to the locations of signals in the data: searching for groups, clusters, or regions with a high proportion of signals, or for a long initial stretch of p-values containing many signals. Each of these methods can be described as searching for a *specific* type of structure and then reweighting the p-values in a *data-adaptive* way. Conversely, an existing method by Genovese et al. [15] handles *general* structures by reweighting the p-values in a *non-adaptive* way. Each p-value $P_i$ is reweighted using prior information (i.e. the reweighting does not depend on the p-values themselves), then the BH procedure is applied to the reweighted p-values $\{P_i/w_i\}$. In this setting, the $w_i$'s can be chosen to reflect any type of known structure, for instance, giving priority to certain hypotheses over others (like in the ordered testing setting) or to certain groups of hypotheses, but the weights cannot be chosen as functions of the p-values themselves.

Our proposed method, which we introduce next, combines these two lines of work by choosing adaptively-chosen weights (that is, weights which may depend on the p-values themselves) that can be designed to reflect any type of structure believed to be present in the problem; our method can handle general structures with data-adaptive weights.

## 2.2 The structure-adaptive Benjamini-Hochberg procedure

We now define our new method, the structure-adaptive Benjamini-Hochberg algorithm.

**Definition 1** (Structure-adaptive Benjamini-Hochberg algorithm (SABHA)). *Given a target FDR level $\alpha \in [0, 1]$, a threshold $\tau \in [0, 1]$, and values $\widehat{q}_1, \ldots, \widehat{q}_n \in [0, 1]$ (where $\widehat{q}_i$ represents an estimated probability that the $i$th test corresponds to a null), define*

$$\widehat{k} = \max\left\{k \geq 1 : P_i \leq \left(\frac{\alpha}{\widehat{q}_i} \cdot \frac{k}{n}\right) \wedge \tau \text{ for at least } k \text{ many p-values } P_i\right\},$$

with the convention that $\widehat{k} = 0$ if this set is empty. Then the SABHA method rejects any p-value $P_i$ satisfying

$$P_i \leq \left( \frac{\alpha}{\widehat{q}_i} \cdot \frac{\widehat{k}}{n} \right) \wedge \tau, \tag{2}$$

for a total of $\widehat{k}$ many rejections.

Note that, if we set $\tau = 1$ and $\widehat{q}_i = 1$ for all $i$, then this is exactly the original Benjamini-Hochberg procedure; if instead we take $\tau \in (0, 1)$ and set $\widehat{q}_i = \widehat{\pi}_0$ to be an estimate of the proportion of nulls (which is constant across all $i$), then this can give us Storey's modification of the BH procedure. Alternately, if we choose $\widehat{q}$ to be a fixed vector (i.e. not dependent on $P$), then this is equivalent to Genovese et al. [15] p-value weighting method, their weights $w_i$ are given by $1/\widehat{q}_i$ in our notation.

To understand our proposed method intuitively, we sketch a coarse estimate for the false discovery proportion of this method. We consider a random model where each p-value $P_i$ has a probability $q_i$ of being a null. Suppose that $\widehat{q}_i \approx q_i$ is an accurate approximation. For some fixed $k$, we would like to know how many nulls satisfy $P_i \leq \left( \frac{\alpha}{\widehat{q}_i} \cdot \frac{k}{n} \right) \wedge \tau$. We have

$$\mathbb{E}\left[ \sum_{i=1}^{n} \mathbb{1}\left\{ i \text{ is null and } P_i \leq \left( \frac{\alpha}{\widehat{q}_i} \cdot \frac{k}{n} \right) \wedge \tau \right\} \right] \approx \sum_{i=1}^{n} q_i \cdot \left[ \left( \frac{\alpha}{q_i} \cdot \frac{k}{n} \right) \wedge \tau \right] \leq \alpha k,$$

where the approximation holds since $\widehat{q}_i \approx q_i$, and if $P_i$ is a null p-value then it should be uniformly distributed. Therefore, when we reject $\widehat{k}$ many p-values, we expect that there are $\approx \alpha \widehat{k}$ many nulls among them, leading to a false discovery proportion that is $\approx \alpha$.

In fact, our theoretical results will show that, even without assuming that $\widehat{q}$ estimates some underlying random model over nulls and signals, we can nonetheless bound the false discovery rate of this method at a level that is not much higher than the target level $\alpha$, provided that $\widehat{q}$ is chosen from some low-complexity class of vectors, with a few additional natural constraints.

For further intuition for the SABHA method, we give an empirical-Bayes motivation for the method in Appendix A.

## 3   False discovery control results

We give two different results controlling the false discoveries that might be selected by our method. We will first consider a setting where the p-values are assumed to be independent, and second, a dependent setting where the p-values come from dependent z-scores or from a Gaussian copula model. In each case, we show that the FDR is bounded only slightly higher than the target level $\alpha$, as long as the set $\mathcal{Q}$ is sufficiently restricted. Throughout, we will consider the set of null hypotheses $\mathcal{H}_0 \subset [n]$ as fixed and of course unknown.

Before stating our results on FDR and FDP, we first examine the role of the set $\mathcal{Q} \subset (0, 1]^n$. We will require that $\widehat{q} = \widehat{q}(P)$ is chosen to satisfy

$$\widehat{q}(P) \in \mathcal{Q}, \text{ and either } \sum_{i=1}^{n} \frac{\mathbb{1}\{P_i > \tau\}}{\widehat{q}_i(1 - \tau)} \leq n \text{ or } \widehat{q}(P) = \mathbf{1}_n, \tag{3}$$

where $\mathbf{1}_n = (1, \ldots, 1) \in \mathbb{R}^n$. To understand the intuition behind this assumption, we again consider the random model for nulls and non-nulls: suppose that $P_i$ is uniformly distributed (i.e. null) with probability $q_i$, and the signal strength is such that the non-null p-values are almost never above the threshold $\tau$. Then we would have

$$\Pr(P_i > \tau) \approx \Pr(P_i > \tau \text{ and } P_i \text{ is a null}) = q_i \cdot (1 - \tau),$$

and therefore,
$$\mathbb{E}\left[\sum_{i=1}^{n} \frac{\mathbb{1}\{P_i > \tau\}}{q_i(1-\tau)}\right] \approx n.$$

Since this sum will typically concentrate strongly around its expectation, we see that the bound in (3) should hold approximately for the true probability vector $q$; requiring that this bound holds for $\widehat{q}$ essentially means that we are making sure that $\widehat{q}$ is not chosen to be too small on average. Of course, in practice we may find a data set where the bound $\sum_{i=1}^{n} \frac{\mathbb{1}\{P_i \geq \tau\}}{q_i(1-\tau)} \leq n$ is impossible to obtain for any $q \in \mathcal{Q}$, even if we minimize the sum by setting $q = \mathbf{1}_n$. This will occur when $\sum_i \mathbb{1}\{P_i > \tau\} > n(1-\tau)$. For example, if all the p-values are uniformly distributed, then $\sum_i \mathbb{1}\{P_i > \tau\} \sim \text{Binomial}(n, 1-\tau)$, which will exceed the bound roughly half of the time. In this setting, we have no evidence to believe that there are many signals present in the data set and it is intuitive to set $\widehat{q}_i = 1$ for all $i$.

For our stronger FDR bound in the independent setting, we will also assume that

$$\widehat{q}(P) \text{ depends on the p-values } P \text{ only through } (P_i \cdot \mathbb{1}\{P_i > \tau\})_i. \tag{4}$$

This requirement, that $\widehat{q} = \widehat{q}(P)$ cannot depend on p-values falling below $\tau$, ensures that these low p-values (i.e. any p-values that we might reject, according to the SABHA method) are still "random" even after we compute $\widehat{q}$. That is, even after choosing adaptive weights $\widehat{q}_i$, it still makes sense to run the (weighted) Benjamini-Hochberg algorithm, which assumes uniformly distributed null p-values.

Next, while assumptions (3) and (4) govern the way in which we can choose weights $\widehat{q}$ from the set $\mathcal{Q}$, how should we choose this set $\mathcal{Q}$ initially? In order to give a result controlling FDR, we need to quantify the complexity of $\mathcal{Q}$ in such a way that we control the extent to which $\mathcal{Q}$ can overfit to the data. To do so, we introduce an additional definition: for any set $\mathcal{A} \subseteq \mathbb{R}^n$, the Rademacher complexity of $\mathcal{A}$ is given by[2]

$$\text{Rad}(\mathcal{A}) = \mathbb{E}\left[\frac{1}{n} \sup_{x \in \mathcal{A}} |\langle x, \xi \rangle|\right],$$

where the expectation is taken over a vector $\xi$ of independent Rademacher variables, $\xi_i \overset{\text{iid}}{\sim} \text{Unif}\{\pm 1\}$. It is known that the Rademacher complexity is closely related to Gaussian width and statistical dimension; see e.g. [4, Lemma 4]. We will apply this notion of complexity to the set

$$\mathcal{Q}_{\text{inv}} := \left\{((q_1)^{-1}, \ldots, (q_n)^{-1}) : q \in \mathcal{Q}\right\}.$$

Since $\mathbf{1}_n \in \mathcal{Q}_{\text{inv}}$ always by assumption, we have

$$\text{Rad}(\mathcal{Q}_{\text{inv}}) \geq \mathbb{E}\left[\frac{1}{n}|\langle \mathbf{1}_n, \xi \rangle|\right] \sim n^{-1/2}.$$

On the other hand, if $\|x\|_\infty$ is bounded for all $x \in \mathcal{Q}_{\text{inv}}$ (i.e. $\mathcal{Q}$ is bounded away from zero), then $\text{Rad}(\mathcal{Q}_{\text{inv}})$ can be bounded by a constant. We can therefore expect that $\mathcal{O}(n^{-1/2}) \leq \text{Rad}(\mathcal{Q}_{\text{inv}}) \leq \mathcal{O}(1)$ for any choice of $\mathcal{Q}$ that is bounded away from zero.

With these assumptions and definitions in place for $\widehat{q}$ and $\mathcal{Q}$, we turn to our results on FDR control, and will see how the complexity measure $\text{Rad}(\cdot)$ controls the tradeoff between flexibility and false discovery control.

### 3.1 FDR control under independence

We first give a result on FDR control under a strong independence assumption:

The null $p$-values, $(P_i)_{i \in \mathcal{H}_0}$, are mutually independent,

and are independent from the non-null p-values $(P_i)_{i \notin \mathcal{H}_0}$. (5)

---

[2]Note that some texts define the Rademacher complexity without the absolute value, or with a scale factor $\frac{2}{n}$ in place of $\frac{1}{n}$.

Since in many applications, null p-values might not be exactly uniformly distributed (for instance the p-values may be discretized), we can relax the uniformity assumption, instead requiring that the null p-values are *super-uniform*, i.e. are no smaller than p-values generated from a uniform distribution:

$$P_i \text{ is super-uniform for all } i \in \mathcal{H}_0, \text{ that is, } \Pr(P_i \leq t) \leq t \text{ for all } t \in [0,1]. \tag{6}$$

We now turn to our first main result, proving finite-sample FDR control for the SABHA procedure under independence.

**Theorem 1.** *Fix a target FDR level $\alpha \in [0,1]$, a threshold $\tau \in (0,1)$, and a set $\mathcal{Q} \subseteq (0,1]^n$ with $\mathbf{1}_n \in \mathcal{Q}$. Suppose that the vector of p-values $P \in [0,1]^n$ has independent and super-uniform nulls as in assumptions* (5) *and* (6)*, and $\widehat{q} = \widehat{q}(P)$ satisfies assumptions* (3) *and* (4)*. Then the false discovery rate of the SABHA procedure, run with parameters $\alpha, \tau, \widehat{q}(P)$ over the p-values $P$, is bounded as*

$$\mathrm{FDR} = \mathbb{E}[\mathrm{FDP}] \leq \alpha \left(1 + \frac{\mathrm{Rad}(\mathcal{Q}_{\mathrm{inv}})}{1-\tau}\right),$$

*where the expectation is taken with respect to the distribution of the p-values.*

Note that, since this result holds for any fixed set of nulls $\mathcal{H}_0 \subseteq [n]$, it would also hold in expectation over any random model for the nulls and non-nulls.

To interpret the results of this theorem, we would typically choose a class $\mathcal{Q}$ which is sufficiently simple to ensure that $\mathrm{Rad}(\mathcal{Q}_{\mathrm{inv}}) \ll 1$; in this type of setting, the FDR bound would be only slightly higher than $\alpha$. $\mathcal{Q}$ should reflect our beliefs about the structure of the problem, for instance, we may believe that the signals appear in clusters among our $n$ hypotheses.

In contrast, if we do not sufficiently constrain $\mathcal{Q}$, then the bound on FDR can become meaningless—for instance, if we take $\mathcal{Q} = [\alpha, 1]^n$ and $\tau = 0.5$, then $\mathrm{Rad}(\mathcal{Q}_{\mathrm{inv}}) \approx \frac{\alpha^{-1}-1}{2}$, and so the upper bound on FDR is $\approx 1$. This is not merely an artifact of the theory. For instance, suppose that $P_i \overset{\mathrm{iid}}{\sim} \mathrm{Unif}[0,1]$ for all $i$, and so there is no signal in the data. If we choose $\widehat{q}_i = 1$ whenever $P_i > \tau$ and $\widehat{q}_i = \alpha$ whenever $P_i \leq \tau$, then whenever $\sum_i \mathbb{1}\{P_i \leq \tau\} \geq n\tau$ (which occurs roughly half the time), the SABHA procedure rejects *all* p-values $P_i \leq \tau$. The source of the problem is that the vector $\widehat{q}$ is drastically overfitting to the binary vector $(\mathbb{1}\{P_i > \tau\})_i$, which is pure noise. By instead choosing $\mathcal{Q}$ with a low Rademacher complexity, we ensure that $\widehat{q}$ cannot overfit too much to this data.

## 3.2   FDP control with dependent p-values

While many results in the literature are proved under an assumption of independent p-values, much attention has also been given to the problem of handling dependent p-values, since in many settings the data gathered for hypotheses $1, 2, \ldots, n$ come from the same set of experiments and are therefore likely to be strongly dependent. While the original Benjamini-Hochberg procedure maintains FDR control as long as this dependence satisfies some positivity assumptions (namely the PRDS condition of Benjamini and Yekutieli [7]), this is no longer the case for the null-proportion-adaptive BH procedure (1) proposed by Storey [27].

We now give a result for the SABHA method, in a more general setting where the p-values may be dependent. In order to do so, we consider a model where the p-values are derived from potentially dependent z-statistics. Specifically, suppose that

$$P_i = 1 - \Phi(Z_i), \quad \text{where} \quad Z \sim \mathcal{N}(\mu, \Sigma) \tag{7}$$

for some mean vector $\mu \in \mathbb{R}^n$ and some covariance matrix $\Sigma \in \mathbb{R}^{n \times n}$ with diagonal entries $\Sigma_{ii} = 1$. Then $P_i$ is super-uniformly distributed (i.e. is a null) whenever $\mu_i \leq 0$—this means that these p-values are computing a one-sided, right-tailed test for each $Z_i$. Of course, we could equivalently compute a one-sided left-tailed test by taking $P_i = \Phi(Z_i)$. For simplicity, our results below will be proved for the one-sided setting only, but an analogous result holds for two-sided tests.

More generally, we can consider a Gaussian copula model for the vector of p-values,[3]

$$P_i = f_i(Z_i), \quad \text{where } Z \sim \mathcal{N}(0, \Sigma) \text{ and } f_i \text{ is monotone non-increasing for each } i, \tag{8}$$

that is, the vector $P$ is obtained by applying monotone marginal transformations to a multivariate Gaussian vector. (In the high-dimensional setting, this model is also known as the nonparanormal model [22].) We can assume that the mean is zero, and that $\Sigma_{ii} = 1$ for all $i$, since shifting and rescaling the $Z_i$'s can simply be absorbed into the functions $f_i$. As before, we require that any null p-value is super-uniformly distributed:

$$\text{For all } i \in \mathcal{H}_0, \ \Pr(P_i \leq t) \leq t \text{ for all } t \in [0, 1]. \tag{9}$$

In this setting, we obtain the following bound on the FDR:

**Theorem 2.** *Fix a target FDR level $\alpha \in [0, 1]$, a threshold $\tau \in (0, 1)$, and a set $\mathcal{Q} \subseteq [\epsilon, 1]^n$ with $\mathbf{1}_n \in \mathcal{Q}$. Suppose that the vector of p-values $P \in [0, 1]^n$ follows a Gaussian copula model (8) with super-uniform null p-values (9), and $\widehat{q} = \widehat{q}(P)$ satisfies assumption (3). Then the false discovery rate of the SABHA procedure, run with parameters $\alpha, \tau, \widehat{q}(P)$ over the p-values $P$, is bounded as*

$$\text{FDR} = \mathbb{E}[\text{FDP}] \leq$$
$$\alpha \left[ 1 + \sqrt{\text{Rad}(\mathcal{Q}_{\text{inv}})} \sqrt{\log(en^2)} \cdot \left( \frac{4}{\sqrt{\epsilon}(1-\tau)} + \frac{4\sqrt[4]{\kappa}}{\sqrt{\alpha c}} \right) + \sqrt{\frac{\log(n)}{n}} \cdot \frac{\sqrt{\kappa}}{\alpha c \sqrt{2}} \right] + \Pr\left(\widehat{k} < c \cdot n\right),$$

*where $\kappa$ is the condition number of the covariance $\Sigma$ in the Gaussian copula model (8).*

Here we again see the critical role of the complexity of $\mathcal{Q}$, as measured by $\text{Rad}(\mathcal{Q}_{\text{inv}})$. Treating $\epsilon, \kappa, c$ as fixed positive constants, we can interpret this result as proving that the FDR is bounded nearly at level $\alpha$, as long as $\Pr\left(\widehat{k} < c \cdot n\right)$ is low and

$$\text{Rad}(\mathcal{Q}_{\text{inv}}) \ll \frac{1}{\sqrt{\log(n)}}.$$

(Compare to our FDR bound Theorem 1 in the independent setting, where for a meaningful bound we only require $\text{Rad}(\mathcal{Q}_{\text{inv}}) \ll 1$). Of course, if the Rademacher complexity $\text{Rad}(\mathcal{Q}_{\text{inv}})$ is much smaller than this bound, then we can afford values of $c$ and $\epsilon$ that is close to zero, that is, we would allow for a very small number of discoveries and for small weights $\widehat{q}_i$. For instance, if $\text{Rad}(\mathcal{Q}_{\text{inv}}) \propto \frac{1}{\sqrt{n}}$ (as is the case for the ordered testing setting, which we will discuss in Section 4.1), we can choose any $c, \epsilon \gg \sqrt{\frac{\log(n)}{n}}$—in particular, this would require that the number of discoveries satisfies $\widehat{k} \gg \sqrt{n \log(n)}$ with large probability, rather than a constant proportion of discoveries.

**Comments on assumptions** While this theorem is proved in the setting of z-statistics, we would expect that FDR is (approximately) controlled across a far broader range of distributions, as long as the dependence structure is again well-conditioned; the Gaussian assumption is simply an artifact of our proof technique, which uses a result from the literature regarding subgaussianity of the Gaussian sign vector, $\text{sign}(Z)$.

On the other hand, as mentioned above, strong dependence across a large proportion of the p-values would no longer enable FDR control, for instance, in an equi-correlated setting where $Z \sim N(0, \Sigma)$, where $\Sigma_{ij} = \rho > 0$ for all $i \neq j$. Indeed, in this type of setting, it is known that even Storey's modification of the BH procedure will no longer control FDR, so the more flexible SABHA method will of course lose FDR control as well—therefore, requiring a bound on condition number $\kappa$ appears to be necessary for the FDR control result to hold. To give an example of a dependence structure where we *do* expect FDR control results to hold, we can consider a scenario where the hypotheses have an inherent spatial or temporal structure, with dependence that is strong locally but decays quickly with distance, allowing for a bounded $\kappa$.

---

[3] To obtain the z-scores model (7) as a special case, we would define $f_i(z) = 1 - \Phi(\mu_i + z)$, since the mean vector has been set to zero in (8).

### 3.2.1 Dependent p-values in Storey's method

As mentioned earlier, the null-proportion-adaptive BH procedure (1) proposed by Storey [27] is not guaranteed to control FDR for dependent p-values in general, even if we assume positive dependence. Instead, to characterize the behavior of this method under weaker forms of dependence, Storey et al. [28, Equation (7)] consider an asymptotic setting where the empirical distribution of the null p-values converges to (super-)uniform,

$$\text{For all } t \in [0,1], \frac{\sum_{i \in \mathcal{H}_0^n} \mathbb{1}\{P_i \leq t\}}{n\pi_0} \to G_0(t) \text{ as } n \to \infty \tag{10}$$

for some cumulative distribution function $G_0$ that is super-uniform (i.e. $G_0(t) \leq t$), and the empirical distribution of the non-nulls converges to some alternate distribution,

$$\text{For all } t \in [0,1], \frac{\sum_{i \notin \mathcal{H}_0^n} \mathbb{1}\{P_i \leq t\}}{n(1-\pi_0)} \to G_1(t) \text{ as } n \to \infty \tag{11}$$

where $G_1(t)$ is the cumulative distribution function corresponding to the distribution of non-null p-values. Here $\pi_0$ is the (asymptotic) proportion of nulls, $|\mathcal{H}_0^n|/n$, while $\mathcal{H}_0^n \subset [n]$ is the set of nulls for a particular value of $n$. Additional assumptions of Storey et al. [28, Theorem 4] effectively require also that the adaptive BH procedure is likely to make a nonzero number of discoveries:

$$\text{For some } c > 0, \Pr\left(\widehat{k} \geq c \cdot n\right) \to 1 \text{ as } n \to \infty. \tag{12}$$

With these assumptions in place, Storey et al. [28, Theorem 4] prove that $\limsup_{n \to \infty} \text{FDR} \leq \alpha$.

In the adaptive weights setting, it is no longer natural to consider an asymptotic regime where $n$ tends to infinity, since the set of possible weights $\mathcal{Q}$ is a subset of $(0,1]^n$ and therefore our problem is intrinsically tied to a particular value of $n$. For example, if $\mathcal{Q}$ is defined relative to a graph structure on $n$ nodes representing the $n$ hypotheses (for instance, testing gene expression levels for $n$ genes, with edges representing known interactions between pairs of genes), it is not clear how to embed this into a sequence of problems with increasing $n$. Therefore, it is difficult to compare our finite-sample results in a dependent setting for SABHA, to the asymptotic regime considered by Storey et al. [28]. However, we note that the asymptotic assumption (12) in their work, appears also in our result Theorem 2 where we assume that $\Pr\left(\widehat{k} < c \cdot n\right)$ is low.

## 4  Applications to specific types of structure

We now consider the application of our main result to various specific structured settings. We will compare to existing work for each setting. For each case we also provide bounds on the Rademacher complexity of $\mathcal{Q}_{\text{inv}}$ for the relevant sets $\mathcal{Q}$. Bounding $\text{Rad}(\mathcal{Q}_{\text{inv}})$ allow us to prove an explicit result for FDR control by applying Theorem 1 under the assumption of independent p-values, or Theorem 2 for p-values obtained from dependent z-scores.

All results in this section are proved in Appendix C.

### 4.1  Ordered structure

In an ordered setting, we might believe that the hypotheses early in the list are more likely to contain true signals than those later in the list.

In this setting, one natural choice for $\widehat{q} = \widehat{q}(P)$ would be to require that this vector is nondecreasing. To avoid degeneracy, we also impose a lower bound $\epsilon > 0$, and define the set

$$\mathcal{Q} = \mathcal{Q}_{\text{ord}} = \{q : \epsilon \leq q_1 \leq \cdots \leq q_n \leq 1\}. \tag{13}$$

We could alternately choose to enforce that $\widehat{q}$ is a "step function" of the form $\widehat{q} = (\epsilon, \ldots, \epsilon, 1, \ldots, 1)$, in which case again $\widehat{q} \in \mathcal{Q}_{\text{ord}}$ (in fact, $\mathcal{Q}_{\text{ord}}$ is the convex hull of such "step functions").

Now we examine the FDR control of the SABHA method for this ordered setting.

**Lemma 1.** *For $\mathcal{Q} = \mathcal{Q}_{\text{ord}}$ as defined above,*

$$\text{Rad}(\mathcal{Q}_{\text{inv}}) \leq \frac{1}{\epsilon\sqrt{n}}.$$

Therefore, under the assumptions of Theorem 1 (in the setting where the null p-values are independent),

$$\text{FDR} \leq \alpha \left(1 + \frac{1}{\sqrt{n}} \cdot \frac{1}{\epsilon(1-\tau)}\right),$$

which is barely higher than the target level $\alpha$ (as long as $\epsilon \gg 1/\sqrt{n}$). Similarly, Theorem 2 gives a meaningful result on FDR for the dependent z-scores setting as long as $\epsilon \gg \sqrt{\frac{\log(n)}{n}}$.

**Related methods**  In this ordered scenario, the existing methods of G'Sell et al. [16], Barber and Candès [1], and Li and Barber [21], mentioned earlier in Section 2.1, propose algorithms for choosing an adaptive cutoff point $k$ and labeling $P_1, \ldots, P_k$ as signals and $P_{k+1}, \ldots, P_n$ as nulls. Alternately, the Selective SeqStep method of [1] selects an adaptive cutoff $k$ and then rejects $P_i$ for $i \leq k$ only if $P_i \leq \tau$ for some predetermined threshold $\tau$. In contrast, the ordered adaptive method proposed here does not have a firm cutoff; instead, the p-values early in the list are given higher priority, but rejections may occur at any point in the list, e.g. if the last p-value $P_n$ is extremely low then it is likely to be rejected even if the overall number of rejections is small. The version of our method that is most comparable to these existing works, is when we choose $\widehat{q}$ to be a "step function", $\widehat{q} = (\epsilon, \ldots, \epsilon, 1, \ldots, 1)$; in this case, when $k$ is the data-adaptive number of $\epsilon$ values in $\widehat{q}$, the first $k$ p-values are given high priority for discovery while the remaining $n - k$ p-values are handled as in the original BH procedure, allowing for strong p-values to be rejected even if they appear late in the list. In this sense, this version of our method can be viewed as a relaxation of the ordered testing methods of [16, 1, 21].

**Choosing $\widehat{q} \in \mathcal{Q}$**  To choose a monotone vector $\widehat{q}$ which satisfies $\epsilon \leq \widehat{q}_1 \leq \cdots \leq \widehat{q}_n \leq 1$ and reflects the patterns observed in the data, one approach would be to consider $\widehat{q}_i(1 - \tau)$ as an estimate of $\Pr(P_i > \tau)$, and then maximize the likelihood of this model:

$$\widehat{q} = \underset{q \in \mathbb{R}^n}{\arg\max} \left\{ \sum_i \mathbb{1}\{P_i > \tau\} \log(q_i(1-\tau)) + \mathbb{1}\{P_i \leq \tau\} \log(1 - q_i(1-\tau)) \right. $$
$$\left. : q \in \mathcal{Q}_{\text{ord}}, \sum_i \frac{\mathbb{1}\{P_i > \tau\}}{q_i(1-\tau)} \leq n \right\}.$$

(If we have $\sum_i \mathbb{1}\{P_i > \tau\} > n(1-\tau)$ then we instead set $\widehat{q} = \mathbf{1}_n$ as required by our assumption (3).) We give an algorithm for solving this convex optimization problem in Appendix D.

Alternately, instead of a likelihood-based approach for choosing a monotone vector, we may alternately wish to consider a simpler construction for $\widehat{q}$, given by a "step function" of the form

$$q^k = (\underbrace{\epsilon, \ldots, \epsilon}_{k \text{ times}}, \underbrace{1, \ldots, 1}_{n - k \text{ times}}) \tag{14}$$

as discussed above. The highest power (i.e. largest number of rejections) will be obtained by taking $k$ as large as possible while still ensuring that the assumption (3) (and, if desired, (4)) hold. In this case, we can simply

set
$$K = \max\left\{k = 1, \ldots, n : \sum_{i=1}^{k} \frac{\mathbb{1}\{P_i > \tau\}}{\epsilon(1-\tau)} + \sum_{i=k+1}^{n} \frac{\mathbb{1}\{P_i > \tau\}}{1-\tau} \leq n\right\}, \quad (15)$$

or set $K = 0$ if this set is empty; we then set $\widehat{q} = q^K$ as in (14), which satisfies (3) by our choice of $K$.

Choosing $\widehat{q}$ to be a step function has some similarity to the recent Adaptive SeqStep method of Lei and Fithian [20], which rejects all $P_i \leq \epsilon$ for $i = 1, \ldots, K$, where $\epsilon$ is fixed while $K$ is determined adaptively by estimating the FDP of this set of discoveries using the indicators $\{P_i > \tau\}$ to estimate the proportion of nulls present.[4] To compare these two methods, Adaptive SeqStep finds an adaptive cutoff $K$ and uses a *fixed* rejection threshold $\epsilon$ for all p-values that appear before the cutoff (i.e. $P_1, \ldots, P_K$), while SABHA with $\widehat{q} = q^K$ also finds an adaptive cutoff $K$ but uses an *adaptive* rejection threshold, and also is able to reject p-values after the cutoff (i.e. $P_i$ for $i > K$) if they are extremely low.

## 4.2   Group structure

Suppose that the set of hypotheses is partitioned into groups, $[n] = \mathcal{G}_1 \cup \cdots \cup \mathcal{G}_d$, with group sizes $n_1 + \cdots + n_d = n$, with the signals likely to appear together in these groups according to some natural structure or prior information. Specifically, we might believe that each group has its own proportion of nulls and non-nulls, and might then estimate $\widehat{q}$ to be constant within each group but allow it to vary across groups. In this setting, we assume that these groups are predetermined—they are independent of the p-values. Here, our goal is to find *individual* signals within each group, with the group structure helping increase our power in this search—we are not aiming to reject entire groups.

We define
$$\mathcal{Q} = \mathcal{Q}_{\text{group}} = \{q : \epsilon \leq q_i \leq 1 \text{ for all } i, \text{ and } q_i = q_j \text{ whenever } i, j \text{ are in the same group}\}. \quad (16)$$

(We again impose a lower bound $\epsilon$ to avoid degeneracy.) For this choice of $\mathcal{Q}$, this is in fact equivalent to Hu et al. [17]'s "Group Benjamini-Hochberg" method for group-wise reweighting, where the proportion of nulls in each group, $\pi_0^{(k)} = \frac{|\mathcal{H}_0 \cap \mathcal{G}_k|}{|\mathcal{G}_k|}$, is estimated and then used to recalibrate the BH procedure. To compare notation, in our setting since we choose $\widehat{q}$ to be group-wise constant, we can write $\widehat{q}_i = \widehat{\pi}_0^{(k)}$ for each $i \in \mathcal{G}_k$, for groups $k = 1, \ldots, d$. Hu et al. [17]'s theoretical results offer exact finite-sample FDR control in the oracle setting where the $\pi_0^{(k)}$'s are known rather than estimated, and asymptotic FDR control otherwise. We will now see that, if the partition of $[n]$ into groups is not too fine, then our main result in Theorem 1 offers a finite-sample FDR guarantee when the $\pi_0^{(k)}$'s are estimated adaptively from the data, rather than known in advance.

**Lemma 2.** *For $\mathcal{Q} = \mathcal{Q}_{\text{group}}$ as defined above, if the groups are of sizes $n_1, \ldots, n_d$, then*
$$\text{Rad}(\mathcal{Q}_{\text{inv}}) \leq \frac{1}{2\epsilon n} \sum_{i=1}^{d} \sqrt{n_i}.$$

Therefore, under the assumptions of Theorem 1 (in particular, assuming that the null p-values are independent),
$$\text{FDR} \leq \alpha \left(1 + \frac{1}{2\epsilon(1-\tau)} \cdot \frac{\sum_{i=1}^{d} \sqrt{n_i}}{n}\right).$$

This yields a meaningful bound on FDR as long $\epsilon \gg \frac{\sum_{i=1}^{d} \sqrt{n_i}}{n}$. For example, if the $d$ groups are each of size $n_i = n/d$, then we require $\epsilon \gg \sqrt{d/n}$ in order to bound FDR near $\alpha$. If we instead apply Theorem 2 for the setting of dependent z-scores, we would instead obtain a meaningful FDR guarantee when $\epsilon \gg \sqrt{\frac{d \log(n)}{n}}$.

---

[4] Our parameters $\epsilon, \tau, K$ are equivalent to their notation $s, \lambda, \widehat{k}$.

In summary, the SABHA method in this setting is an example of the procedure proposed by Hu et al. [17] (where the estimated proportion of nulls within each group can be determined with any desired data-adaptive method). Hu et al. [17]'s theoretical results for the adaptive setting prove asymptotic FDR control, while our theoretical results are able to give a strong finite-sample guarantee for the data-adaptive procedure.

**Choosing $\widehat{q} \in \mathcal{Q}$**  To choose a group-wise constant vector $\widehat{q}$, we simply need to choose a value $\widetilde{q}_k$ for each group $k = 1, \ldots, d$ and then set $\widehat{q}_i = \widetilde{q}_k$ for each $i \in \mathcal{G}_k$. Here $\widetilde{q}_k$ should represent our estimated proportion of nulls in group $\mathcal{G}_k$. We will maximize the likelihood subject to the constraints that $\widehat{q}$ is group-wise constant and lies in the range $[\epsilon, 1]$. We take

$$\widetilde{q} = \arg\max_{q \in \mathbb{R}^d} \left\{ \sum_k \left( \sum_{i \in \mathcal{G}_k} \mathbb{1}\{P_i > \tau\} \right) \log(q_k(1-\tau)) + \left( \sum_{i \in \mathcal{G}_k} \mathbb{1}\{P_i \leq \tau\} \right) \log(1 - q_k(1-\tau)) \right.$$
$$\left. : \epsilon \leq q_k \leq 1 \text{ for } k = 1, \ldots, d; \sum_k \frac{\sum_{i \in \mathcal{G}_k} \mathbb{1}\{P_i > \tau\}}{q_k(1-\tau)} \leq n \right\},$$

and then define $\widehat{q} \in \mathbb{R}^n$ accordingly. (If we have $\sum_i \mathbb{1}\{P_i > \tau\} > n(1-\tau)$ then we simply set $\widehat{q} = \mathbf{1}_n$, as always.)

Note that, if each group satisfies

$$\frac{\sum_{i \in \mathcal{G}_k} \mathbb{1}\{P_i > \tau\}}{n_k(1-\tau)} \in [\epsilon, 1] \tag{17}$$

(that is, no group appears to have a proportion of nulls that is $> 1$ or $< \epsilon$), then the constrained maximum likelihood estimator is obtained by simply setting

$$\widetilde{q}_k = \frac{\sum_{i \in \mathcal{G}_k} \mathbb{1}\{P_i > \tau\}}{n_k(1-\tau)}$$

for each group $k = 1, \ldots, d$. This is equivalent to Storey [27]'s method for estimating the proportion of nulls in a list of p-values, except applied separately to each group. When (17) is not satisfied, though, the optimization problem must be solved jointly over all groups; in Appendix D, we give an algorithm for this problem.

### 4.2.1  Special case: adapting to the overall proportion of nulls

As a special case, we can consider the setting where there is only a single group, $\mathcal{G}_1 = [n]$. In this setting, we would then choose $\widehat{q}$ to be some multiple of $\mathbf{1}_n$, the vector of all 1's. Letting $\widehat{\pi}_0 = \widehat{q}_i$ (which is constant across all $i$), SABHA is then equivalent to the Benjamini-Hochberg procedure at threshold $\frac{\alpha}{\widehat{\pi}_0}$ instead of $\alpha$, where our method would calculate

$$\widehat{\pi}_0 = \frac{\sum_i \mathbb{1}\{P_i > \tau\}}{n(1-\tau)},$$

truncated to the interval $[\epsilon, 1]$ if necessary. This coincides exactly with Storey [27]'s modification of the Benjamini-Hochberg procedure, and our result Lemma 2 shows that we should expect the FDR to be no larger than $\alpha \cdot (1 + \mathcal{O}(1/\sqrt{n}))$, very close to the target level $\alpha$. (Storey et al. [28] prove *exact* FDR control for this adaptive procedure, but in their setting they take a slightly more conservative definition of $\widehat{\pi}_0$.)

### 4.2.2  Special case: positive and negative effects

As a second special case, we can consider incorporating sign information into our tests. Suppose that our p-values are computed from real-valued statistics $X_1, \ldots, X_n$, whose null distribution is continuous and symmetric. Letting $F_0$ be the cumulative distribution function for the null, for a two-sided test we would set $P_i = 2(1 - F_0(|X_i|))$. Now we define two groups, $\mathcal{G}_+ = \{i : X_i > 0\}$ and $\mathcal{G}_- = \{i : X_i < 0\}$. We then estimate the proportion of nulls separately for each of these two groups. The effect of separating into these two groups is that we gain power if the true signals are more likely to be positive than negative (or vice versa), as compared to estimating a single $\widehat{\pi}_0$ for the entire set of $n$ hypotheses. This version of our procedure is equivalent to Zhao and Fung [30]'s procedure discussed in Section 2.1.

Of course, this appears to contradict our requirement that the groups are chosen ahead of time, i.e. independently of the p-values. In fact, in some settings, this does not pose a problem. The reason is that by assuming that the null $X_i$'s are independent, and each have a symmetric distribution (as is the case for z-scores), we ensure that the null $|X_i|$'s are independent from $\text{sign}(X_i)$ (and furthermore from the other signs, $\text{sign}(X_j)$ for $j \neq i$). Therefore, we can determine the groups $\mathcal{G}_+, \mathcal{G}_-$ without revealing any information about the null p-values $P_i$, which are a function of $|X_i|$ only and do not depend on $\text{sign}(X_i)$ when using a two-sided test.

### 4.3    Low total variation

In some settings, the $n$ hypotheses may exhibit some form of locally smooth or locally constant structure. For instance, if each hypothesis is associated with a spatial location, then it might be natural to assume that signals are spatially clustered, meaning that nearby hypotheses have equal or similar likelihoods of being null or non-null. We might also have this sort of local similarity in other settings, for instance, similarity of individuals within a data set.

To generalize this setting, consider a connected undirected graph $G = (V_G, E_G)$ on nodes $V_G = [n]$ with $e_G = |E_G|$ many undirected edges. We will search for a vector $\widehat{q}$ that is locally constant on this graph, meaning that $q_i - q_j = 0$ for most edges $(i,j) \in E_G$.

One choice for $\mathcal{Q}$ is given by

$$\mathcal{Q} = \mathcal{Q}_{\text{TV-sparse}} = \left\{ q \in [\epsilon, 1]^n : \sum_{(i,j) \in E} \mathbb{1}\{q_i \neq q_j\} \leq m \right\}, \tag{18}$$

where $\epsilon > 0$ bounds the values away from zero to avoid degeneracy, and $m$ is some predetermined bound on the number of non-constant edges in the graph. However, $\mathcal{Q}_{\text{TV-sparse}}$ is not convex, and it may be difficult to choose a $\widehat{q}$ in this set. For this reason we may wish to consider a convex set,

$$\mathcal{Q} = \mathcal{Q}_{\text{TV-}\ell_1} = \left\{ q \in [\epsilon, 1]^n : \sum_{(i,j) \in E} |q_i - q_j| \leq m \right\}, \tag{19}$$

which is a strict relaxation of the set considered above—that is, $\mathcal{Q}_{\text{TV-sparse}} \subseteq \mathcal{Q}_{\text{TV-}\ell_1}$—since for $q \in [0,1]^n$,

$$\sum_{(i,j) \in E} |q_i - q_j| \leq \sum_{(i,j) \in E} \mathbb{1}\{q_i \neq q_j\}.$$

In either case, the parameter $m$ is a tuning parameter that should be specified in advance, perhaps reflecting prior knowledge about the amount of total variation in the underlying structure of the data; our theory will treat $m$ as fixed rather than data-adaptive.

We will bound the relevant Rademacher complexities by making use of recent work by Hütter and Rigollet [18]. First, following [18], define the incidence matrix of the graph $G$, $D_G \in \{-1, 0, +1\}^{e_G \times n}$, which for each edge $(i,j) \in E_G$ has a corresponding row with $i$th entry $+1$, $j$th entry $-1$, and zeros elsewhere; define also the quantity

$$\rho_G = \max_{k=1,\ldots,e_G} \|(D_G^+)_k\|_2,$$

where $D_G^+ \in \mathbb{R}^{n \times e_G}$ is the pseudo-inverse of $D_G$ and $(D_G^+)_k$ is its $k$th column. Our next result shows that $\rho_G$ controls the complexity of the set of adaptive weights $\mathcal{Q}$.

**Lemma 3.** For $\mathcal{Q} = \mathcal{Q}_{\text{TV-sparse}}$ as defined above,

$$\text{Rad}(\mathcal{Q}_{\text{inv}}) \leq \frac{1}{\epsilon\sqrt{n}} + \frac{2\rho_G m \sqrt{\log(n)}}{\epsilon n}.$$

If we instead take $\mathcal{Q} = \mathcal{Q}_{\text{TV-}\ell_1}$, then

$$\text{Rad}(\mathcal{Q}_{\text{inv}}) \leq \frac{1}{\epsilon\sqrt{n}} + \frac{2\rho_G m \sqrt{\log(n)}}{\epsilon^2 n}.$$

Therefore, if we assume independent p-values, then applying Theorem 1, we obtain

$$\text{FDR} \leq \alpha \left(1 + \frac{1}{\epsilon(1-\tau)\sqrt{n}} + \frac{2\rho_G m \sqrt{\log(n)}}{(\epsilon \text{ or } \epsilon^2) \cdot (1-\tau)n}\right)$$

with power of $\epsilon$ in the second denominator given by $\epsilon$ for $\mathcal{Q} = \mathcal{Q}_{\text{TV-sparse}}$ or $\epsilon^2$ for $\mathcal{Q} = \mathcal{Q}_{\text{TV-}\ell_1}$. To see an example of the resulting scaling, consider a two-dimensional grid on an array of $\sqrt{n} \times \sqrt{n}$ nodes, for which it is shown in Hütter and Rigollet [18, Proposition 4] that $\rho_G = \mathcal{O}\left(\sqrt{\log(n)}\right)$. Therefore, we obtain nontrivial FDR guarantees for $\epsilon \gg \max\left\{\frac{1}{\sqrt{n}}, \frac{m \log(n)}{n}\right\}$ in the total variation sparsity setting ($\mathcal{Q} = \mathcal{Q}_{\text{TV-sparse}}$), and for $\epsilon \gg \sqrt{\frac{m \log(n)}{n}}$ in the total variation norm setting ($\mathcal{Q} = \mathcal{Q}_{\text{TV-}\ell_1}$). For the one-dimensional case, where $G$ is a chain graph on $n$ nodes, Hütter and Rigollet [18, Section 3.1] calculate $\rho_G = \sqrt{n}$, and so to obtain a nontrivial FDR guarantee via Theorem 1 with $\mathcal{Q} = \mathcal{Q}_{\text{TV-sparse}}$ we need $\epsilon \gg \sqrt{\frac{m^2 \log(n)}{n}}$, or choosing $\mathcal{Q} = \mathcal{Q}_{\text{TV-}\ell_1}$ we instead need $\epsilon \gg \sqrt[4]{\frac{m^2 \log(n)}{n}}$. For simplicity, we do not give the results implied by Theorem 2 for the dependent z-scores setting here, but these are also straightforward to calculate.

**Comparison to related methods**   To compare to methods such as those of Chouldechova [11] and Siegmund et al. [25], which search for signals that appear in clusters (in a spatial domain or related setting), their work aims to make discoveries that form clusters (determined by the data) and to measure error at the cluster-wise level as well. In contrast, our work sets the weights $\widehat{q}_i$ in a locally constant way so that the *threshold* for making a discovery is locally constant, but the p-values themselves may still lead to discoveries which do not form contiguous or well-defined clusters. In general, for our method, the pattern of discoveries that we might expect to see, would show some regions with high proportions of discoveries, and other regions where discoveries are very scarce. This gap would typically much wider than if we were to apply the BH procedure to the same data, since SABHA boosts our ability to make discoveries in high-signal regions.

**Choosing $\widehat{q} \in \mathcal{Q}$**   As before, we can consider a constrained maximum likelihood approach to choosing $\widehat{q} \in \mathcal{Q}$, for either $\mathcal{Q} = \mathcal{Q}_{\text{TV-sparse}}$ or $\mathcal{Q} = \mathcal{Q}_{\text{TV-}\ell_1}$:

$$\widehat{q} = \underset{q \in \mathbb{R}^n}{\arg\max} \left\{ \sum_i \mathbb{1}\{P_i > \tau\} \log(q_i(1-\tau)) + \mathbb{1}\{P_i \leq \tau\} \log(1 - q_i(1-\tau)) \right.$$
$$\left. : q \in \mathcal{Q}, \sum_i \frac{\mathbb{1}\{P_i > \tau\}}{q_i(1-\tau)} \leq n \right\}. \quad (20)$$

(If we have $\sum_i \mathbb{1}\{P_i > \tau\} > n(1-\tau)$ then we instead set $\widehat{q} = \mathbf{1}_n$, as always.) If we use $\mathcal{Q} = \mathcal{Q}_{\text{TV-}\ell_1}$ then this is a convex optimization problem; we give an algorithm for this setting in Appendix D. If instead we use $\mathcal{Q} = \mathcal{Q}_{\text{TV-sparse}}$ this is a highly nonconvex problem and may be very difficult to solve.

# 5 Proof of false discovery control

In this section we prove our main results on false discovery control, Theorem 1 and Theorem 2.

## 5.1 Complexity of a set: Rademacher complexity and cube complexity

We first develop a result on set complexity, which we will use in the proof of Theorem 1. For any set $\mathcal{A} \subseteq \mathbb{R}^n$, recall that the Rademacher complexity of $\mathcal{A}$ is defined as

$$\text{Rad}(\mathcal{A}) = \mathbb{E}\left[\frac{1}{n} \sup_{x \in \mathcal{A}} |\langle x, \xi \rangle|\right] \text{ where } \xi_i \stackrel{\text{iid}}{\sim} \text{Unif}\{\pm 1\}.$$

We now define the "cube complexity", which is similar to the Rademacher complexity but uses arbitrary mean-zero product distributions on the cube $[-1, 1]^n$. First define $\mathfrak{C}$ to be the set of mean-zero product distributions on $[-1, 1]^n$, that is, each $\mathcal{D} \in \mathfrak{C}$ is a distribution of the form $\mathcal{D}_1 \times \cdots \times \mathcal{D}_n$ where each $\mathcal{D}_i$ is a mean-zero distribution on $[-1, 1]$. Then define

$$\text{CubeComplexity}(\mathcal{A}) = \sup_{\mathcal{D} \in \mathfrak{C}} \mathbb{E}_{Y \sim \mathcal{D}}\left[\frac{1}{n} \sup_{x \in \mathcal{A}} |\langle x, Y \rangle|\right].$$

Note that, if we take each $\mathcal{D}_i$ to be the distribution placing probability 0.5 on $+1$ and on $-1$, then the $Y_i$'s are independent Rademacher variables, i.e. $Y$ has the same distribution as $\xi$ above, and then we would have $\mathbb{E}_{Y \sim \mathcal{D}}\left[\frac{1}{n} \sup_{x \in \mathcal{A}} |\langle x, Y \rangle|\right] = \text{Rad}(\mathcal{A})$; this proves that $\text{CubeComplexity}(\mathcal{A}) \geq \text{Rad}(\mathcal{A})$. We now prove that the cube complexity is in fact equal to the Rademacher complexity—that is, the supremum over $\mathcal{D} \in \mathfrak{C}$ is attained by taking $\mathcal{D}_i = \text{Unif}\{\pm 1\}$ for each $i$.

**Lemma 4.** *For any set $\mathcal{A} \subseteq \mathbb{R}^n$, the cube complexity satisfies* $\text{CubeComplexity}(\mathcal{A}) = \text{Rad}(\mathcal{A})$.

*Proof of Lemma 4.* We know that $\text{CubeComplexity}(\mathcal{A}) \geq \text{Rad}(\mathcal{A})$ trivially by choosing $\mathcal{D}_i = \text{Unif}\{\pm 1\}$ for each $i$. Now we prove the reverse bound. Fix any $\mathcal{D} \in \mathfrak{C}$ and let $Y \sim \mathcal{D}$. Next, let $U_i \stackrel{\text{iid}}{\sim} \text{Unif}[0, 1]$ be drawn independently of $Y$, and define $\xi \in \{\pm 1\}^n$ as

$$\xi_i = \begin{cases} +1, & U_i \leq \frac{1+Y_i}{2}, \\ -1, & U_i > \frac{1+Y_i}{2}. \end{cases}$$

Then we see that

$$\mathbb{E}\left[\xi_i \mid Y\right] = 1 \cdot \Pr\left(U_i \leq \frac{1+Y_i}{2} \;\middle|\; Y\right) + (-1) \cdot \Pr\left(U_i > \frac{1+Y_i}{2} \;\middle|\; Y\right) = \frac{1+Y_i}{2} - \left(1 - \frac{1+Y_i}{2}\right) = Y_i,$$

and marginally, the $\xi_i$'s are independent with

$$\mathbb{E}\left[\xi_i\right] = \mathbb{E}\left[\mathbb{E}\left[\xi_i \mid Y\right]\right] = \mathbb{E}\left[Y_i\right] = 0.$$

In other words, after marginalizing over $Y$, the $\xi_i$'s are independent Rademacher variables.

We then have

$$\mathbb{E}\left[\frac{1}{n}\bigg| \sup_{x \in \mathcal{A}} \langle x, Y \rangle \bigg|\right] = \mathbb{E}\left[\frac{1}{n} \sup_{x \in \mathcal{A}} |\langle x, \mathbb{E}\left[\xi \mid Y\right] \rangle|\right] \leq \mathbb{E}\left[\mathbb{E}\left[\frac{1}{n} \sup_{x \in \mathcal{A}} |\langle x, \xi \rangle| \;\middle|\; Y\right]\right] \quad \text{by Jensen's inequality}$$

$$= \mathbb{E}\left[\frac{1}{n} \sup_{x \in \mathcal{A}} |\langle x, \xi \rangle|\right] \quad \text{by the tower law of expectations}$$

$$= \text{Rad}(\mathcal{A}) \quad \text{by definition of Rademacher complexity.}$$

Since this holds for any choice $\mathcal{D} \in \mathfrak{C}$, this proves the desired bound, $\text{CubeComplexity}(\mathcal{A}) \leq \text{Rad}(\mathcal{A})$. □

## 5.2 Proof of FDR control (Theorems 1 and 2)

We now turn to proving our FDR control results for SABHA.

First we work with the expression for the false discovery proportion, to construct an upper bound on FDP that consists of several key terms. We will then bound each term in expectation, under the independence model (Theorem 1) and under the dependent z-statistics model (Theorem 2).

$$\text{FDP} = \sum_{i \in \mathcal{H}_0} \frac{\mathbb{1}\{P_i \text{ is rejected}\}}{1 \vee \widehat{k}} = \sum_{i \in \mathcal{H}_0} \frac{\mathbb{1}\left\{P_i \leq \frac{\alpha \cdot \widehat{k}}{\widehat{q}_i \cdot n} \wedge \tau\right\}}{1 \vee \widehat{k}} \quad \text{by definition of the SABHA method}$$

$$= \sum_{i \in \mathcal{H}_0} \frac{\alpha}{\widehat{q}_i \cdot n} + \sum_{i \in \mathcal{H}_0} \frac{\mathbb{1}\left\{P_i \leq \frac{\alpha \cdot \widehat{k}}{\widehat{q}_i \cdot n} \wedge \tau\right\} - \frac{\alpha \cdot (1 \vee \widehat{k})}{\widehat{q}_i \cdot n}}{1 \vee \widehat{k}}$$

$$= \alpha \cdot \left[1 + \underbrace{\left(\sum_{i \in \mathcal{H}_0} \frac{1}{\widehat{q}_i \cdot n}\right) - 1}_{\text{Term 1}} + \underbrace{\sum_{i \in \mathcal{H}_0} \frac{\mathbb{1}\left\{P_i \leq \frac{\alpha \cdot \widehat{k}}{\widehat{q}_i \cdot n} \wedge \tau\right\} - \frac{\alpha \cdot (1 \vee \widehat{k})}{\widehat{q}_i \cdot n}}{\alpha \cdot (1 \vee \widehat{k})}}_{\text{Term 2}}\right],$$

where in the last two steps we are simply rearranging terms. Examining Term 1 more closely, we recall our assumption (3) on the choice of weights $\widehat{q}$: either $\widehat{q} = \mathbf{1}_n$, or $\widehat{q}$ satisfies $\sum_i \frac{\mathbb{1}\{P_i > \tau\}}{(1-\tau)n \cdot \widehat{q}_i} \leq 1$. In the first case, we trivially have Term $1 \leq 0$ (since $|\mathcal{H}_0| \leq n$), while in the second case,

$$\text{Term 1} = \left(\sum_{i \in \mathcal{H}_0} \frac{1}{\widehat{q}_i \cdot n}\right) - 1 \leq \sum_{i \in \mathcal{H}_0} \frac{1}{\widehat{q}_i \cdot n} - \sum_i \frac{\mathbb{1}\{P_i > \tau\}}{(1-\tau)n \cdot \widehat{q}_i} \leq \sum_{i \in \mathcal{H}_0} \frac{1 - \frac{\mathbb{1}\{P_i > \tau\}}{1-\tau}}{\widehat{q}_i \cdot n}$$

$$\leq \sup_{q \in \mathcal{Q}} \sum_{i \in \mathcal{H}_0} \frac{1 - \frac{\mathbb{1}\{P_i > \tau\}}{1-\tau}}{q_i \cdot n}.$$

Therefore, we can rewrite the above as

$$\text{FDP} \leq \alpha \cdot \left[1 + \underbrace{\max\left\{0, \sup_{q \in \mathcal{Q}} \sum_{i \in \mathcal{H}_0} \frac{1 - \frac{\mathbb{1}\{P_i > \tau\}}{1-\tau}}{q_i \cdot n}\right\}}_{\text{Term 1}} + \underbrace{\sum_{i \in \mathcal{H}_0} \frac{\mathbb{1}\left\{P_i \leq \frac{\alpha \cdot \widehat{k}}{\widehat{q}_i \cdot n}\right\} - \frac{\alpha \cdot (1 \vee \widehat{k})}{\widehat{q}_i \cdot n}}{\alpha \cdot (1 \vee \widehat{k})}}_{\text{Term 2}}\right].$$

Therefore, if both Term 1 and Term 2 are fairly small, then FDP will be not much larger than $\alpha$. Since we are aiming to bound the FDR in our main results, we will only need to bound Term 1 and Term 2 in expectation.

### 5.2.1 Bounding Term 1 and Term 2 under independence

We can bound Term 1 trivially: defining

$$Y_i = \begin{cases} \Pr(P_i > \tau) - \mathbb{1}\{P_i > \tau\}, & i \in \mathcal{H}_0 \\ 0, & i \notin \mathcal{H}_0, \end{cases}$$

we see that $Y \in [-1, 1]^n$, with independent and mean-zero coordinates. Therefore,

$$\mathbb{E}\left[\frac{1}{n} \sup_{q \in \mathcal{Q}} \left|\sum_{i \in \mathcal{H}_0} Y_i / q_i\right|\right] = \mathbb{E}\left[\frac{1}{n} \sup_{x \in \mathcal{Q}_{\text{inv}}} |\langle x, Y \rangle|\right] \leq \text{CubeComplexity}(\mathcal{Q}_{\text{inv}}) = \text{Rad}(\mathcal{Q}_{\text{inv}}),$$

by Lemma 4. Furthermore, $\Pr(P_i > \tau) \geq 1 - \tau$ for all $i \in \mathcal{H}_0$, which proves that

$$\mathbb{E}[\text{Term 1}] \leq \frac{1}{1-\tau} \cdot \mathbb{E}\left[\frac{1}{n} \sup_{q \in \mathcal{Q}} \left|\sum_{i \in \mathcal{H}_0} Y_i/q_i\right|\right] \leq \frac{1}{1-\tau} \cdot \text{Rad}(\mathcal{Q}_{\text{inv}}).$$

Next we turn to Term 2, and will show that in fact $\mathbb{E}[\text{Term 2}] \leq 0$.

Fix any $i \in \mathcal{H}_0$, and define $k_i \in \{1, \ldots, n\}$ to be the output of the SABHA procedure when $P$ is replaced with

$$P_{i \to 0} = (P_1, \ldots, P_{i-1}, 0, P_{i+1}, \ldots, P_n).$$

Note that, by our assumption (4) on the choice of $\widehat{q}$, if $P_i \leq \tau$ then we have $\widehat{q}(P) = \widehat{q}(P_{i \to 0})$. As is often used in proofs for the BH procedure (see e.g. Ferreira and Zwinderman [14] for this type of proof technique), we observe that if $P_i$ is rejected, then replacing it with a smaller p-value would not change the outcome of the procedure—that is, if $P_i$ is rejected, then $\widehat{k} = k_i$. By definition of SABHA, we can rewrite this as

$$P_i \leq \left(\frac{\alpha}{\widehat{q}_i} \cdot \frac{\widehat{k}}{n}\right) \wedge \tau \;\Rightarrow\; P_i \text{ rejected} \;\Rightarrow\; \widehat{k} = k_i. \tag{21}$$

Now, conditioning on $P_{i \to 0}$, we have

$$\Pr\left(P_i \leq \left(\frac{\alpha}{\widehat{q}_i} \cdot \frac{k_i}{n}\right) \wedge \tau \;\Big|\; P_{i \to 0}\right) \leq \left(\frac{\alpha}{\widehat{q}_i} \cdot \frac{k_i}{n}\right) \wedge \tau \leq \frac{\alpha}{\widehat{q}_i} \cdot \frac{k_i}{n}, \tag{22}$$

while the inequality holds due to the super-uniformity and independence assumptions, (5) and (6). Therefore, we can write

$$\mathbb{E}\left[\frac{\mathbb{1}\left\{P_i \leq \frac{\alpha \cdot \widehat{k}}{\widehat{q}_i \cdot n} \wedge \tau\right\}}{1 \vee \widehat{k}} - \frac{\alpha}{\widehat{q}_i \cdot n} \;\Big|\; P_{i \to 0}\right] = \mathbb{E}\left[\frac{\mathbb{1}\left\{P_i \leq \frac{\alpha \cdot k_i}{\widehat{q}_i \cdot n} \wedge \tau\right\}}{k_i} - \frac{\alpha}{\widehat{q}_i \cdot n} \;\Big|\; P_{i \to 0}\right] \leq 0,$$

where the first step holds since, if the indicator variable in the numerator is nonzero, then (21) proves that $\widehat{k} = k_i \geq 1$, while the second step holds by (22) (noting that $k_i$ is fixed, once we condition on $P_{i \to 0}$). Marginalizing over $P_{i \to 0}$, and summing over all $i \in \mathcal{H}_0$, we obtain $\mathbb{E}[\text{Term 2}] \leq 0$, as desired.

Combining our bounds on Term 1 and Term 2, we have proved Theorem 1 for FDR control under independence.

### 5.2.2 Bounding Term 1 and Term 2 for dependent z-scores

The bounds on Term 1 and Term 2 in the dependent z-scores setting are more technical. Since the p-values are no longer independent, we cannot hope to obtain $\mathbb{E}[\text{Term 2}] \leq 0$; instead, we will upper bound this term as

$$\mathbb{E}\left[\text{Term 2} \cdot \mathbb{1}\left\{\widehat{k} \geq c \cdot n\right\}\right] \leq \mathbb{E}\left[\sup_{q \in \mathcal{Q}; k \in \{c \cdot n, \ldots, n\}} \sum_{i \in \mathcal{H}_0} \frac{\mathbb{1}\left\{P_i \leq \frac{\alpha \cdot k}{q_i \cdot n}\right\} - \frac{\alpha \cdot k}{q_i \cdot n}}{\alpha \cdot k}\right],$$

and use the assumption that $\Pr\left(\widehat{k} < c \cdot n\right)$ is low. The key steps of our proof are:

1. First, we bound the covering number of $\mathcal{Q}_{\text{inv}}$ with respect to the $\ell_\infty$ norm, in terms of the Rademacher complexity $\text{Rad}(\mathcal{Q}_{\text{inv}})$, using bounds calculated in Srebro et al. [26]. This allows us to write Term 1 and Term 2 as a maximum over finitely many $q$'s, rather than a supremum over all $q \in \mathcal{Q}$.

2. Next, we apply a result from Barber and Kolar [2], proving that, for a multivariate Gaussian vector $Z$, the centered sign vector $\text{sign}(Z) - \mathbb{E}[\text{sign}(Z)]$ is subgaussian. Since Term 1 and Term 2 are both functions of indicator variables $\mathbb{1}\{P_i \leq c_i\}$ (or, equivalently, $\mathbb{1}\{P_i > c_i\}$) for various thresholds $c_i$, using the Gaussian copula model for $P$ allows us to transform Term 1 and Term 2 into functions of the sign vector $\text{sign}(Z)$, so that we can use the subgaussianity result.

We defer the details of this proof to Appendix B.

# 6  Experiments

We now present empirical results on simulated and real data. For the simulated data, we will examine the low total variation setting described in Section 4.3, and in particular, will test the role of the total variation constraint on the power and FDR of the resulting method. Then, we will demonstrate applications of the other two types of structure described earlier on real data: ordered structure (as described in Section 4.1) with gene/drug response data, and grouped structure (as described in Section 4.2) with fMRI data.[5]

## 6.1  Simulated data: low total variation

In our simulated data experiments, the signals are arranged in a two-dimensional grid, with low total variation in the underlying probabilities of each hypothesis being null. Our estimate of $\widehat{q}$ uses the convex total variation norm as in (19) over the two-dimensional grid. To generate the data, we create a list of $n = 225$ p-values arranged in a $15 \times 15$ grid, and assign a "true" underlying prior probability of each p-value being a null,

$$q_i = \begin{cases} 0.1, & \text{if } i \text{ is in the high-signal region } R, \\ 0.9, & \text{if } i \notin R, \end{cases}$$

where the region $R$ consists of two quarter-circles, in the top-right and bottom-left corner of the grid, for a total of $|R| = 70$ points (see Figure 1). The mean probability of a null is $\frac{1}{n}\sum_i q_i = 0.6511$. To generate the p-values themselves, we draw $I_i \sim \text{Bernoulli}(1-q_i)$ indicating whether hypothesis $i$ is a true signal, and then draw $Z_i \stackrel{\perp\!\!\!\perp}{\sim} N(\mu_i, 1)$ where $\mu_i = \mu_{\text{sig}} \cdot I_i$, for some fixed $\mu_{\text{sig}} > 0$ (with larger $\mu_{\text{sig}}$ indicating a stronger signal). Then we run two-sided z-tests, $P_i = 2(1 - \Phi(|Z_i|))$, where $\Phi$ is the CDF of the standard normal. We repeat our experiment for each value $\mu_{\text{sig}} \in \{0.5, 1, 1.5, \ldots, 3.5\}$.

We implement SABHA on the two-dimensional grid graph, choosing $\tau = 0.5$ and defining $\mathcal{Q} = \mathcal{Q}_{\text{TV-}\ell_2}$ as in (19) with the lower bound parameter set at $\epsilon = 0.1$ and with the total variation constraint $m \in \{10, 15, 20\}$. (Note that, for the "true" $q$, we have $\|q\|_{\text{TV-}\ell_1} = 22.2$, that is, $q \notin \mathcal{Q}$ for any of these values of $m$, but nonetheless these values are sufficient to allow for substantial adaptivity.) We then compare SABHA with BH, and also with Storey's modification of BH ("Storey-BH") given in (1), implemented with parameter $\tau = 0.5$.

To fit $\widehat{q}$ for the SABHA method, i.e. to solve the optimization problem (20), details are given in Appendix D. For comparison we also run an "oracle" version of SABHA where $\widehat{q} = q$, the true vector of prior probabilities of each p-value being a null. For all methods we choose the target FDR level $\alpha = 0.1$.

We then compare the observed FDR and power for each of the considered methods. As we can see in the left panel of Figure 1, for all methods, the average power and average observed FDR both increase with larger $\mu$ (stronger signal). The BH, Storey-BH, and oracle SABHA methods all control FDR at level $\alpha = 0.1$; the BH is in fact known to control FDR at the level $\alpha \cdot \frac{|\mathcal{H}_0|}{n}$, and is therefore quite conservative (i.e. FDR $< 0.1$) as we observe in our results, while the other two methods have FDR $\approx 0.1$. For SABHA, as the signal grows stronger, the observed FDR grows very close to 0.1 for small $m = 10$ and medium $m = 15$; for large $m = 20$, the observed FDR somewhat exceeds the target level $\alpha = 0.1$, as $\widehat{q}$ is overfitting to the p-values. To compare the power, BH is most conservative (lowest power) followed by Storey-BH. Oracle SABHA shows the highest power, while SABHA shows power increasing as the total variation constraint $m$ increases. Here, SABHA is consistently more powerful than BH and Storey-BH over the range of $\mu_{\text{sig}}$.

---
[5]Code for all experiments is available at http://www.stat.uchicago.edu/~rina/sabha.html.

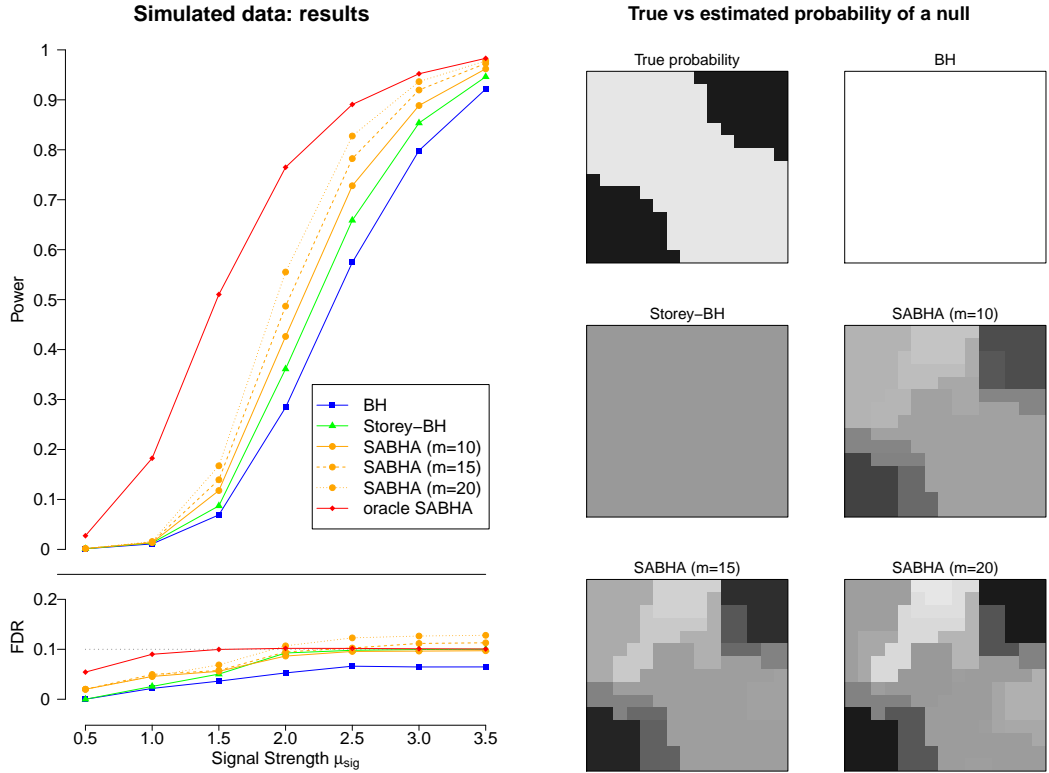

Figure 1: Left: power and observed FDR level of all procedures averaged over 50 trials; the target FDR level is $\alpha = 0.1$. Right: true $q$ vs. estimated $\widehat{q}$ for a single trial with $\mu_{\text{sig}} = 2.5$ (black = 0 = contains all signals, white = 1 = contains all nulls). See Section 6.1 for details.

In the right panel of Figure 1, we further compare the methods according to their estimated probabilities of a null, for one trial at signal strength $\mu_{\text{sig}} = 2.5$. For SABHA with $m = 10, 15, 20$, the plot displays the estimated $\widehat{q}$. We can compare against the true $q$, which is also the input to the oracle SABHA method. Finally, since the BH and Storey-BH methods are equivalent to taking $\widehat{q} = \mathbf{1}_n$ and $\widehat{q} = \widehat{\pi}_0 \cdot \mathbf{1}_n$ and proceeding as with SABHA, we display these estimated $\widehat{q}$'s as well. As expected, for SABHA, we see that the fitted weights $\widehat{q}_i$ approximate the overall pattern of underlying probabilities $q_i$, and the larger values of $m$ show more overfitting in the sense that the weights $\widehat{q}_i$ show distinctions between regions of the grid, which are not actually different in the underlying model (leading to the increased FDR observed previously).

## 6.2 Gene/drug response data: ordered structure

We next apply SABHA to an ordered testing problem, the gene dosage response application described in Li and Barber [21]. The data [12] consists of $n = 22283$ genes' expression levels at various drug dosages.[6] For each gene $i = 1, \ldots, 22283$, we calculate a p-value $P_i$ which evaluates evidence for a change in gene expres-sion level between the control (no dose) setting and a low dose setting, using a permutation test. The genes are arranged in order, with the beginning of the list (low index $i$) corresponding to genes whose differential expression level at *high* dosage leads us to believe that this gene is more likely to exhibit a nonzero response at *low* dosage; for the low dosage data we then run a one-sided test, for instance, a gene that showed increased expression at the high dosage is tested for increased expression at the low dosage. See [21] for more details on the data set and on how the ordering and the p-values are calculated.

---

[6]Data available at http://www.ncbi.nlm.nih.gov/sites/GDSbrowser?acc=GDS2324 or via the GEOquery package [13] in R.

We run SABHA using $\mathcal{Q} = \mathcal{Q}_{\text{ord}}$ (13) and choose $\widehat{q}$ to be a step function as in (14), with lower bound parameter $\epsilon = 0.1$ and threshold $\tau = 0.5$. We compare SABHA against BH, and against Storey-BH with threshold $\tau = 0.5$. We also compare against three methods for ordered hypothesis testing: the ForwardStop method [16], the SeqStep method [1] (with parameter $C = 2$), the accumulation test method with the Hinge-Exp function [21] (parameter $C = 2$), and the Adaptive SeqStep method [20]; see Section 4.1 for some more details on this family of methods. We run each method with target FDR level $\alpha = 0.01, 0.02, \ldots, 0.5$.

In this data set, the sample size (number of measurements for each gene) at each of the three settings—control, low dose, and high dose—is 5. We repeat our experiment three times: first, using this full sample size; second, using a sample size of 2 for the high dose setting only, so as to reduce the quality of the ordering (that is, we will be less accurate in our assessment of which genes are most likely to show a change in expression level at the low dosage); and third, using a random ordering (that is, there is no information contained in the ordering; true signals are equally likely to appear anywhere on the list).

**Results** Figure 2 shows the results from all the methods across the range of $\alpha$ values, for all three settings for how the ordering is determined. The plotted outcome is the number of "discoveries", i.e. the number of genes selected as showing a significant difference between the low dose and no dose setting. Both the BH and Storey-BH methods, which do not use the information of the ordering, are not able to make more than a few discoveries at any level $\alpha$ below $\approx 0.3$; since these methods do not use the ordering information, their outcomes are identical across the three settings. Turning to the ordered testing methods (accumulation test/HingeExp, SeqStep, ForwardStop, Adaptive SeqStep), we see that these methods are able to recover a substantial number of genes even for low $\alpha$ when the ordering is highly informative, with a wide range of performance across the three methods, but make essentially no discoveries when the ordering is random (carries no information).

Next, we see that SABHA is able to perform well across the range of scenarios; while it's not the optimal method in any single scenario, it is the only method whose performance is adaptive while the existing methods are all specialized to one or the other extreme. For the first setting (sample size 5 used to obtain a highly informative ordering ordering), we see that SABHA is less powerful than the best ordered testing method but nonetheless gives strong performance even at low $\alpha$ values; for the second setting (sample size 2 used to obtain a moderately informative ordering), SABHA is now more comparable to the best ordered testing method across much of the range of $\alpha$ values; and for the third setting (a random i.e. completely uninformative ordering), SABHA continues to perform well, and is nearly as powerful as Storey-BH which makes the most discoveries in this setting, while the ordered testing methods now have effectively zero power except for Adaptive SeqStep at the high values of $\alpha$. To summarize, SABHA is able to adapt to the amount of information carried in the ordering, achieving good performance relative to the best method in each setting. In practice, then, when we do not know ahead of time whether the ordered structure is informative or not, SABHA is a good choice as it can adapt to any scenarios.

### 6.3 fMRI data: grouped structure

We now show an application of our method in an analysis of fMRI data, in which the group structure of the problem can be exploited (see Section 4.2 for more details on the group structured setting). The data, gathered by Keller et al. [19] and available online,[7] consists of fMRI measurements taken for subjects who are shown two stimuli in a sequence, a picture and a sentence in either order, with the task of determining whether the two agree. We restrict our attention to data from a single individual (Subject 04867), and to the 20 trials with the picture displayed first and the sentence second, where the procedure is of the form

$$\underbrace{\text{Picture displayed (4 sec)} \ ; \ \text{No stimulus (4 sec)}}_{\text{Picture phase (8 sec)}} \ ; \ \underbrace{\text{Sentence displayed (4 sec)} \ ; \ \text{No stimulus (4 sec)}}_{\text{Sentence phase (8 sec)}}.$$

---

[7]Data available at http://www.cs.cmu.edu/afs/cs.cmu.edu/project/theo-81/www/

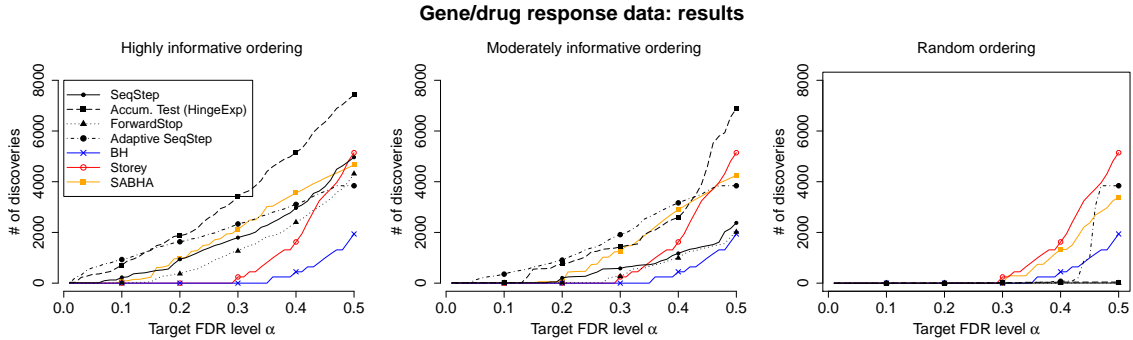

Figure 2: Results for the differential gene expression experiment: for each method, the plot shows the number of discoveries made (i.e. the number of genes selected as showing significant change in expression at the low drug dosage as compared to the control), at a range of target FDR values $\alpha$.

Measurements are taken every 0.5 seconds, for a total of 16 brain images each in the picture phase and in the sentence phase. Though only a fraction of the brain of each subject was imaged, these images reflect 3D activity information, as each images includes 8 two-dimensional slices. The 4691 measured voxels are grouped into 24 regions of interest (ROIs), anatomically or functionally distinct regions in the brain.[8]

We are interested in how the activity of different parts of the brain differs between the picture phase and the sentence phase, which can be formulated as a multiple testing problem, with each voxel in the brain corresponding to a hypothesis. For each voxel we compute the average activity level across the 16 images in each of the two phases, for each of the 20 trials, which leads to a p-value computed from a paired t-test with sample size 20. In the first few columns of Figure 3, we display the partition into ROIs (Figure 3(a)), the original data averaged for the picture phase and the sentence phase (Figure 3(b)), and the calculated p-values (Figure 3(c)). Each column is a single 3D image of the brain, split into eight horizontal slices.

We then implement the SABHA method with group structure determined by the ROIs; we use $\mathcal{Q} = \mathcal{Q}_{\text{group}}$ as in (16), and choose parameters $\tau = 0.5$ and $\epsilon = 0.1$. (Recall that, for the group structure setting, this method is an example of the Group Benjamini-Hochberg method proposed by Hu et al. [17], where the proportion of nulls within each group can be estimated with any data-adaptive method; here we specifically use our proposed estimate $\widehat{q}$ given in (16).) We compare against BH and against Storey-BH with threshold $\tau = 0.5$. For all methods we use target FDR level $\alpha = 0.2$.

**Results**   The results from the fMRI experiment are displayed in Figure 3. The number of discoveries made by each method is:

| Method | # discoveries |
|---|---|
| BH | 931 |
| Storey-BH | 1217 |
| SABHA | 1234 |

---

[8]While 4698 voxels are measured and 25 ROIs are defined, 7 voxels are not labeled with a ROI, and one ROI is not assigned to any voxels, so we remove these from the data.

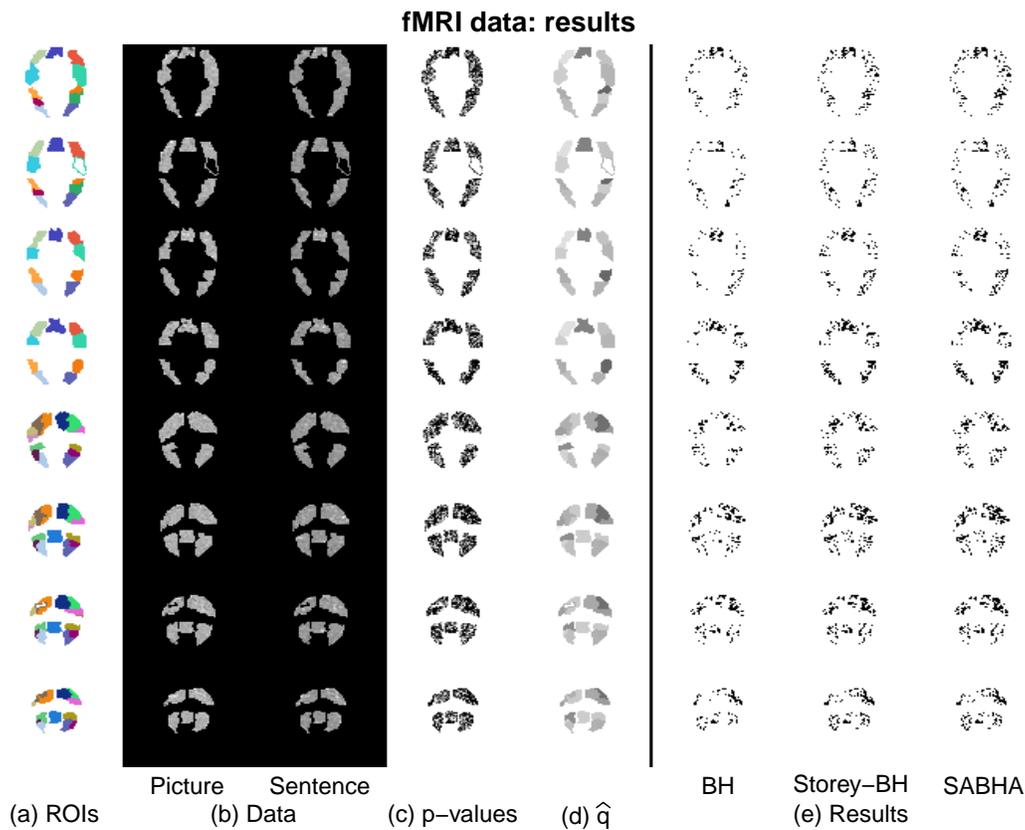

Figure 3: Results for the fMRI data; the eight images in each column are the horizontal slices of the brain. (a) The 24 ROIs defined in the data set, each pictured in a different color. (b) The average activity levels recorded in the experiment for the picture phase and for the sentence phase (white = highest activity level, black = zero activity level). (c) The p-values obtained at each voxel using the paired t-test (black = 0 = most significant, white = 1 = least significant). (d) The estimated vector $\widehat{q}$ for SABHA using the group structure defined by the ROIs (black = 0 = contains all signals, white = 1 = contains all nulls). (e) Results for the BH, Storey-BH, and SABHA methods (a black point indicates a voxel labeled as a discovery).

In Figure 3(d) we see the estimated $\widehat{q}$ for the SABHA method, and Figure 3(e) displays the locations of the discoveries for each of the three methods.

To understand the performance of SABHA in greater detail, consider the estimated $\widehat{q}$ in Figure 3(d); we see that some of the ROIs are estimated to have a much lower proportion of nulls (those ROIs that appear darker in the figure), and these are the regions that show the greatest gains in the number of discoveries made by SABHA as compared to the other methods. To see the difference more quantitatively, in Figure 4 we display the proportion of discoveries made in each ROI by each of the three methods, compared to the estimated value for $\widehat{q}$ in this ROI. As expected, the greatest gains for SABHA are in those ROIs where $\widehat{q}$ is estimated to be lowest. In contrast, in ROIs where $\widehat{q}$ is estimated to be near 1, Storey-BH makes more discoveries. This is because Storey-BH effectively estimates a uniform (constant) $\widehat{q}$ across all ROIs (i.e. $\widehat{q} = \widehat{\pi}_0 \mathbf{1}_n$ where $\widehat{\pi}_0$ is the *overall* estimated proportion of nulls), while for SABHA this estimated proportion varies across ROIs, and will thus be lower for some ROIs and higher for others. Overall, using the group structure allows for more discoveries—and may perhaps be more accurate as it uses information that is more locally relevant for each ROI, although of course we cannot assess this without knowing the "ground truth".

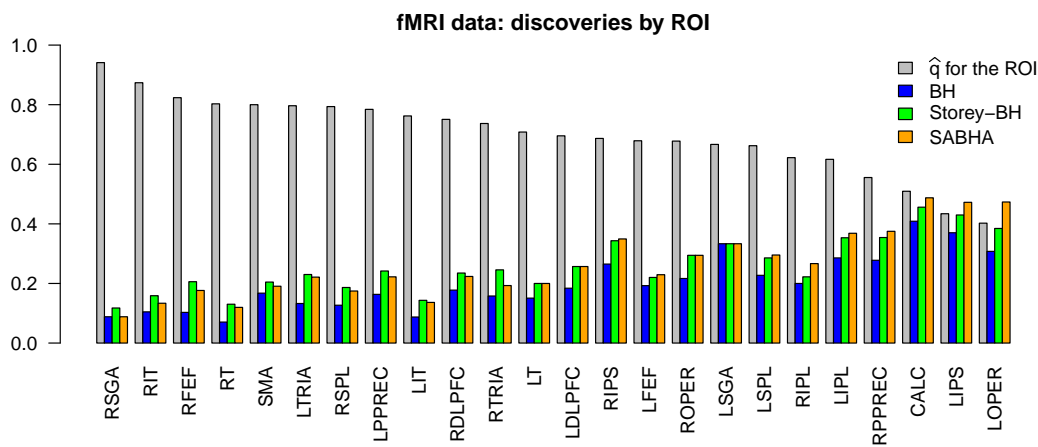

Figure 4: Proportion of discoveries within each ROI for the BH, Storey-BH, and SABHA methods, compared to the estimated proportion of nulls, $\widehat{q}$. The ROIs are sorted in decreasing order of the estimated $\widehat{q}$.

# 7 Discussion

Exploiting the structure within multiple testing problems can lead to substantially increased power. In this paper, we have proposed SABHA, a unified approach to structured multiple testing which is adaptive to diverse structures in the patterns of signals and nulls among the hypotheses. We have demonstrated SABHA in three simulated and real data examples, each representing a commonly observed type of structure: ordered, grouped, and low total variation. SABHA leads to higher power than methods that do not exploit the known structure, while maintaining control of the FDR as long as the adaptive weights are constrained sufficiently so as not to excessively overfit to the data. Our results cover both the independent setting (that is, p-values are independent), and a dependent setting under some assumptions on the empirical distribution of the p-values. Though in this work we have considered three specific types of structures, the SABHA framework is general and can be customized to adapt to any desired type of structure in the pattern of signals and nulls, allowing for greater power in a potentially wide range of applications where multiple testing problems arise.

#### Acknowledgements

This work was partially supported by an Alfred P. Sloan Fellowship and by NSF award DMS-1654076.

# A    An empirical-Bayes interpretation

To understand the SABHA procedure through the lens of empirical Bayesian methods, we first review the empirical Bayes interpretation of the BH procedure and of Storey's modification (following Benjamini and Hochberg [6, Section 3.3] and Storey [27, Section 4]). Consider a two-groups model, where each hypothesis is independently assigned to be null (with probability $\pi_0$) or non-null (with probability $1-\pi_0$). Then, the null p-values are drawn from the uniform distribution, while the non-null p-values are drawn from some alternate distribution $F_1$. To summarize, the p-values are drawn independently from a mixture model,

$$P_i \stackrel{\text{iid}}{\sim} \pi_0 \cdot \text{Unif}[0,1] + (1-\pi_0) \cdot F_1 =: F.$$

Now, if we were to choose a rejection threshold $t$, and reject all p-values satisfying $P_i \leq t$, then our *Bayesian* false discovery rate is given by

$$\text{BayesFDR}(t) = \Pr(P_i \text{ is null} \mid P_i \leq t) = \frac{\Pr(P_i \text{ is null and } \leq t)}{\Pr(P_i \leq t)} = \frac{\pi_0 \cdot t}{\pi_0 \cdot t + (1-\pi_0) \cdot F_1(t)} = \frac{\pi_0 \cdot t}{F(t)}. \tag{23}$$

We can then estimate the mixture distribution $F$ empirically,

$$F(t) \approx \widehat{F}_n(t) = \frac{1}{n} \sum_i \mathbb{1}\{P_i \leq t\}.$$

Plugging in $\widehat{F}_n(t)$ for $F(t)$ in the denominator, and some estimate $\widehat{\pi}_0$ for the null proportion $\pi_0$ in the numerator, we then have

$$\text{BayesFDR}(t) = \frac{\pi_0 \cdot t}{F(t)} \approx \frac{\widehat{\pi}_0 \cdot n \cdot t}{\sum_i \mathbb{1}\{P_i \leq t\}}.$$

(Of course, if we instead use the upper bound $\pi_0 \leq 1$, then we would obtain an approximate upper bound $\text{BayesFDR}(t) \lessapprox \frac{n \cdot t}{\sum_i \mathbb{1}\{P_i \leq t\}}$.) We can then choose the largest threshold $\widehat{t}$ such that our approximation of the Bayes FDR is bounded by the target level $\alpha$, and reject all p-values $P_i \leq \widehat{t}$. It is known that this procedure is equivalent to Storey's modification of the BH method (or, if $\pi_0 \leq 1$ is used in place of the approximation $\widehat{\pi}_0$, to the BH method itself); this equivalence can be seen by taking $\widehat{t} = \frac{\alpha \widehat{k}}{n}$, where $\widehat{k}$ is the number of rejections, as defined in Section 2.1.

We now consider a more general setting, where the $n$ hypotheses are no longer assumed to be exchangeable—some hypotheses may be much more likely than others to contain a true signal (i.e. $\pi_0$ may not be the same for every $i$), and some signals may be stronger than others (i.e. the distribution $F_1$ of the non-null p-values may not be the same for every $i$). To formalize this, suppose that the p-values are drawn as

$$P_i \stackrel{\perp\!\!\!\perp}{\sim} q_i \cdot \text{Unif}[0,1] + (1-q_i) \cdot (F_1)_i =: F_i,$$

where now the probability of being null, and the signal strength for the non-nulls, may differ across the $n$ p-values. Of course, with non-identically-distributed p-values, we may want the thresholds to be different for each $i$ as well—calculating the Bayes FDR (i.e. the posterior probability of being null) for each $i$ individually, we would like to choose threshold $t_i$ such that

$$\alpha = \Pr(P_i \text{ is null} \mid P_i \leq t_i) = \frac{q_i \cdot t_i}{q_i \cdot t_i + (1-q_i) \cdot (F_1)_i(t_i)} = \frac{q_i \cdot t_i}{F_i(t_i)}. \tag{24}$$

Of course, we are no longer able to estimate this denominator—in contrast to the i.i.d. setting, where we have access to $n$ draws from the distribution $F$ (and so $F(t) \approx \widehat{F}_n(t)$ is a very accurate approximation), now we only observe a single draw $P_i$ from each distribution $F_i$. This lack of information seems insurmountable.

However, since we are only aiming to control the FDR, these posterior probabilities only need to be accurately estimated *on average* over all $n$ p-values. With this in mind, we replace (24) with the following expression,

$$\alpha = \frac{q_i \cdot n \cdot t_i}{\sum_{i'} F_{i'}(t_{i'})}, \tag{25}$$

where we have averaged the denominators. Of course, this new denominator can indeed be approximated:

$$\sum_{i'} F_{i'}(t_{i'}) = \sum_{i'} \Pr\left(P_{i'} \leq t_{i'}\right) \approx \sum_{i'} \mathbb{1}\left\{P_{i'} \leq t_{i'}\right\},$$

which is in fact equal to the number of rejections at our chosen thresholds $t_1, \ldots, t_n$. Returning to (25), we plug in estimates $\widehat{q}_i$ as well as our estimated denominator, and solve for the threshold $t_i$:

$$\alpha = \frac{q_i \cdot n \cdot t_i}{\sum_{i'} F_{i'}(t_{i'})} \approx \frac{\widehat{q}_i \cdot n \cdot t_i}{\# \text{ rejections}} \Rightarrow t_i = \frac{\alpha}{\widehat{q}_i} \cdot \frac{\# \text{ rejections}}{n},$$

which exactly agrees with the definition of the SABHA method, given in (2), aside from the truncation at $\tau$.

# B   Proof of Theorem 2

In this section, we fill in the details for the proof of Theorem 2. Recall from Section 5.2 that

$$\text{FDP} \leq \alpha \cdot \left[ 1 + \underbrace{\max\left\{0, \sup_{q \in \mathcal{Q}} \sum_{i \in \mathcal{H}_0} \frac{1 - \frac{\mathbb{1}\{P_i > \tau\}}{1-\tau}}{q_i \cdot n}\right\}}_{\text{Term 1}} + \underbrace{\sum_{i \in \mathcal{H}_0} \frac{\mathbb{1}\left\{P_i \leq \frac{\alpha \cdot \widehat{k}}{\widehat{q}_i \cdot n}\right\} - \frac{\alpha \cdot (1 \vee \widehat{k})}{\widehat{q}_i \cdot n}}{\alpha \cdot (1 \vee \widehat{k})}}_{\text{Term 2}} \right].$$

Therefore, we now need to bound Term 1 and Term 2 in this setting.

## B.1   Covering number

First, we write $\mathcal{N}_\infty(\mathcal{Q}_{\text{inv}} \cup \{0\}, \delta)$ to denote the covering number of $\mathcal{Q}_{\text{inv}} \cup \{0\}$ with respect to the $\ell_\infty$ norm at scale $\delta$, that is, the smallest cardinality of a set $\mathcal{A}_\delta \subset \mathbb{R}^n$ such that, for all $x \in \mathcal{Q}_{\text{inv}} \cup \{0\}$, there is some $y \in \mathcal{A}_\delta$ with $\|x - y\|_\infty \leq \delta$. The following lemma bounds this covering number:

**Lemma 5** (Adapted from Srebro et al. [26, Lemmas A.2 and A.3]). *For any $\delta > 0$ and any $\mathcal{B} \subset \mathbb{R}^n$,*

$$\sqrt{\log\left(\mathcal{N}_\infty(\mathcal{B}, \delta)\right)} \leq \frac{2\text{Rad}(\mathcal{B})\sqrt{n \log(en^2)}}{\delta}.$$

Noting that $\text{Rad}(\mathcal{Q}_{\text{inv}}) = \text{Rad}(\mathcal{Q}_{\text{inv}} \cup \{0\})$, this means that we can assume

$$\sqrt{\log(|\mathcal{A}_\delta|)} \leq \frac{2\text{Rad}(\mathcal{Q}_{\text{inv}})\sqrt{n \log(en^2)}}{\delta}.$$

Next, by replacing each $y \in \mathcal{A}_\delta$ with $y + \delta \cdot \mathbf{1}_n$, we can instead guarantee that, for each $x \in \mathcal{Q}_{\text{inv}} \cup \{0\}$, we have $x \leq y \leq x + 2\delta \cdot \mathbf{1}_n$ (where the bounds hold elementwise). Since $\mathcal{Q}_{\text{inv}} \cup \{0\} \subset [0, \epsilon^{-1}]^n$, without loss of generality we can further assume that $\mathcal{A}_\delta \subset [0, \epsilon^{-1}]^n$ also, by projecting each $y \in \mathcal{A}_\delta$ to this set. From this point on, we treat this set $\mathcal{A}_\delta$ as fixed.

## B.2 Bounding Term 1

First, we bound

$$\text{Term 1} = \max\left\{0, \sup_{q \in \mathcal{Q}} \sum_{i \in \mathcal{H}_0} \frac{1 - \frac{\mathbb{1}\{P_i > \tau\}}{1-\tau}}{q_i \cdot n}\right\} \leq \frac{1}{n} \sup_{x \in \mathcal{Q}_{\text{inv}} \cup \{0\}} \langle x, Y \rangle,$$

where we recall that $\Pr(P_i > \tau) \geq 1 - \tau$ for all nulls $i \in \mathcal{H}_0$, and define $Y$ as the random vector with entries

$$Y_i = \begin{cases} 1 - \frac{\mathbb{1}\{P_i > \tau\}}{\Pr(P_i > \tau)}, & i \in \mathcal{H}_0, \\ 0, & i \notin \mathcal{H}_0. \end{cases}$$

By definition of $\mathcal{A}_\delta$, for any $x \in \mathcal{Q}_{\text{inv}} \cup \{0\}$, we can find some $y \in \mathcal{A}_\delta$ so that $x \leq y \leq x + 2\delta \cdot \mathbf{1}_n$ holds elementwise. Then

$$\langle x, Y \rangle = \langle y, Y \rangle + \langle x - y, Y \rangle \leq \langle y, Y \rangle + n \cdot \|x - y\|_\infty \cdot \|Y\|_\infty \leq \langle y, Y \rangle + \frac{2\delta n}{1 - \tau},$$

and so

$$\text{Term 1} \leq \frac{2\delta}{1-\tau} + \frac{1}{n} \max_{y \in \mathcal{A}_\delta} \langle y, Y \rangle.$$

Next, for each $i$, define $t_i = \sup\{t : f_i(t) > \tau\}$, so that $f_i(t) > \tau$ for all $t < t_i$ and $f_i(t) \leq \tau$ for all $t > t_i$, since the marginal transformation $f_i$ is assumed to be non-increasing. Since $\Pr(P_i > \tau) \geq 1 - \tau$ for any null p-value, we have $\Phi(t_i) = \Pr(Z_i \leq t_i) = \Pr(P_i > \tau) \geq 1 - \tau$ for all $i \in \mathcal{H}_0$. Then we can rewrite

$$Y_i = \begin{cases} 1 - \frac{\mathbb{1}\{Z_i \leq t_i\}}{\Phi(t_i)}, & i \in \mathcal{H}_0, \\ 0, & i \notin \mathcal{H}_0. \end{cases}$$

which may be incorrect on an event of probability zero (i.e. $Z_i = t_i$ for some $i$), but we can ignore this possibility since we are only working with expected values. We can rewrite this again as

$$Y_i = a_i \cdot (\text{sign}(Z_i - t_i) - \mathbb{E}[\text{sign}(Z_i - t_i)]), \text{ where } a_i = \begin{cases} \frac{1}{2\Phi(t_i)} \leq \frac{1}{2(1-\tau)}, & i \in \mathcal{H}_0, \\ 0, & i \notin \mathcal{H}_0. \end{cases}$$

So, for each $y \in \mathcal{A}_\delta$, we have

$$\langle y, Y \rangle = \langle y, a \circ (\text{sign}(Z - t) - \mathbb{E}[\text{sign}(Z - t)]) \rangle = \langle a \circ y, \text{sign}(Z - t) - \mathbb{E}[\text{sign}(Z - t)] \rangle,$$

where $\circ$ denotes elementwise product, $t$ is the vector with entries $t_i$, and $\text{sign}(Z - t)$ is taken elementwise. Barber and Kolar [2, Lemma 4.5] proves that the vector $\text{sign}(Z - t) - \mathbb{E}[\text{sign}(Z - t)]$ is $\kappa$-subgaussian, meaning that $\langle y, Y \rangle$ is $\kappa \cdot \|a \circ y\|_2^2$-subgaussian. We calculate $\|a \circ y\|_2^2 \leq n\|a\|_\infty^2 \|y\|_\infty^2 \leq \frac{n}{4(1-\tau)^2 \epsilon^2}$. Therefore,

$$\mathbb{E}\left[\max_{y \in \mathcal{A}_\delta} \langle y, Y \rangle\right] \leq \sqrt{2\log(|\mathcal{A}_\delta|)} \cdot \frac{\sqrt{n}}{2(1-\tau)\epsilon}.$$

Combining everything, and plugging in our bound on $|\mathcal{A}_\delta|$, we obtain

$$\mathbb{E}[\text{Term 1}] \leq \frac{2\delta}{1-\tau} + \frac{1}{n} \cdot \sqrt{2\log(|\mathcal{A}_\delta|)} \cdot \frac{\sqrt{n}}{2(1-\tau)\epsilon} \leq \frac{2\delta}{1-\tau} + \frac{\text{Rad}(\mathcal{Q}_{\text{inv}})\sqrt{2\log(en^2)}}{\delta\epsilon(1-\tau)}.$$

Finally, we set $\delta = \sqrt{\frac{\text{Rad}(\mathcal{Q}_{\text{inv}})\sqrt{\log(en^2)}}{\epsilon\sqrt{2}}}$ to obtain

$$\mathbb{E}[\text{Term 1}] \leq \frac{4}{1-\tau}\sqrt{\frac{\text{Rad}(\mathcal{Q}_{\text{inv}})\sqrt{\log(en^2)}}{\epsilon}}.$$

## B.3  Bounding Term 2

Next, assuming that $\widehat{k} \geq c \cdot n$, we bound

$$\text{Term 2} = \sum_{i \in \mathcal{H}_0} \frac{\mathbb{1}\left\{P_i \leq \frac{\alpha \cdot \widehat{k}}{\widehat{q}_i \cdot n}\right\} - \frac{\alpha \cdot (1 \vee \widehat{k})}{\widehat{q}_i \cdot n}}{\alpha \cdot (1 \vee \widehat{k})} \leq \sup_{x \in \mathcal{Q}_{\text{inv}}, k \in \{c \cdot n, \ldots, n\}} \sum_{i \in \mathcal{H}_0} \frac{\mathbb{1}\left\{P_i \leq \frac{\alpha \cdot k}{n} \cdot x_i\right\} - \frac{\alpha \cdot k}{n} \cdot x_i}{\alpha \cdot k}.$$

By definition of $\mathcal{A}_\delta$, for any $x \in \mathcal{Q}_{\text{inv}} \cup \{0\}$, we can find some $y \in \mathcal{A}_\delta$ so that $x \leq y \leq x + 2\delta \cdot \mathbf{1}_n$ holds elementwise, and so

$$\sum_{i \in \mathcal{H}_0} \frac{\mathbb{1}\left\{P_i \leq \frac{\alpha \cdot k}{n} \cdot x_i\right\} - \frac{\alpha \cdot k}{n} \cdot x_i}{\alpha \cdot k} \leq \sum_{i \in \mathcal{H}_0} \frac{\mathbb{1}\left\{P_i \leq \frac{\alpha \cdot k}{n} \cdot y_i\right\} - \frac{\alpha \cdot k}{n} \cdot (y_i - 2\delta)}{\alpha \cdot k}$$

$$\leq \sum_{i \in \mathcal{H}_0} \frac{\mathbb{1}\left\{P_i \leq \frac{\alpha \cdot k}{n} \cdot y_i\right\} - \frac{\alpha \cdot k}{n} \cdot y_i}{\alpha \cdot k} + 2\delta,$$

where the last step holds since $|\mathcal{H}_0| \leq n$. So, we have

$$\text{Term 2} \leq 2\delta + \max_{y \in \mathcal{A}_\delta; k \in \{c \cdot n, \ldots, n\}} \sum_{i \in \mathcal{H}_0} \frac{\mathbb{1}\left\{P_i \leq \frac{\alpha \cdot k}{n} \cdot y_i\right\} - \frac{\alpha \cdot k}{n} \cdot y_i}{\alpha \cdot k}. \tag{26}$$

Now, for each $y$ and each $k$, define

$$(t_{y,k})_i = \inf\left\{t : f_i(t) \leq \frac{\alpha \cdot k}{n} \cdot y_i\right\},$$

so that $f_i(t) > \frac{\alpha \cdot k}{n}$ for all $t < t_i$ and $f_i(t) \leq \frac{\alpha \cdot k}{n}$ for all $t > t_i$, since the marginal transformation $f_i$ is assumed to be non-increasing. Since $\Pr\left(P_i \leq \frac{\alpha \cdot k}{n}\right) \leq \frac{\alpha \cdot k}{n}$ for any null p-value, we have $1 - \Phi(t_i) = \Pr\left(Z_i > t_i\right) = \Pr\left(P_i \leq \frac{\alpha \cdot k}{n}\right) \leq \frac{\alpha \cdot k}{n}$ for all $i \in \mathcal{H}_0$. We can rewrite (26) as

$$\text{Term 2} \leq 2\delta + \max_{y \in \mathcal{A}_\delta; k \in \{c \cdot n, \ldots, n\}} \sum_{i \in \mathcal{H}_0} \frac{\mathbb{1}\{Z_i > (t_{y,k})_i\} - \Pr(Z_i > (t_{y,k})_i)}{\alpha \cdot k}$$

$$= 2\delta + \max_{y \in \mathcal{A}_\delta; k \in \{c \cdot n, \ldots, n\}} \sum_{i \in \mathcal{H}_0} \frac{\text{sign}(Z_i - (t_{y,k})_i) - \mathbb{E}\left[\text{sign}(Z_i - (t_{y,k})_i)\right]}{2\alpha \cdot k},$$

where these calculations may be incorrect on a set of measure zero (i.e. if $Z_i = (t_{y,k})_i$ exactly, for some $i, y, k$), but we can ignore this as we are only concerned with expected values. We can rewrite this again as

$$\text{Term 2} \leq 2\delta + \max_{y \in \mathcal{A}_\delta; k \in \{c \cdot n, \ldots, n\}} \left\langle \frac{1}{2\alpha k} \mathbf{1}_{\mathcal{H}_0}, \text{sign}(Z - t_{y,k}) - \mathbb{E}\left[\text{sign}(Z - t_{y,k})\right]\right\rangle,$$

where $\mathbf{1}_{\mathcal{H}_0}$ is the vector with $i$th entry equal to $\mathbb{1}\{i \in \mathcal{H}_0\}$, and $\text{sign}(Z - t_{y,k})$ is taken elementwise. By Barber and Kolar [2, Lemma 4.5], the vector $\text{sign}(Z - t_{y,k}) - \mathbb{E}\left[\text{sign}(Z - t_{y,k})\right]$ is $\kappa$-subgaussian, and so the inner product $\langle \frac{1}{2\alpha k} \mathbf{1}_{\mathcal{H}_0}, \text{sign}(Z - t_{y,k}) - \mathbb{E}\left[\text{sign}(Z - t_{y,k})\right]\rangle$ is subgaussian with constant $\kappa \cdot \|\frac{1}{2\alpha k} \mathbf{1}_{\mathcal{H}_0}\|_2^2 \leq \frac{\kappa n}{4\alpha^2 (c \cdot n)^2} = \frac{\kappa}{4\alpha^2 c^2 n}$. Therefore, recalling that we have assumed $\widehat{k} \geq c \cdot n$ in order to obtain this bound,

$$\mathbb{E}\left[\text{Term 2} \cdot \mathbb{1}\left\{\widehat{k} \geq c \cdot n\right\}\right] \leq 2\delta + \sqrt{2 \log(n \cdot |\mathcal{A}_\delta|)} \cdot \frac{\sqrt{\kappa}}{2\alpha c \sqrt{n}}$$

$$\leq 2\delta + \sqrt{2 \log(n) + 2\left(\frac{2\text{Rad}(\mathcal{Q}_{\text{inv}})\sqrt{n \log(en^2)}}{\delta}\right)^2} \cdot \frac{\sqrt{\kappa}}{2\alpha c \sqrt{n}},$$

by plugging in our bound on $|A_\delta|$. Finally, setting $\delta = \sqrt{\frac{\mathrm{Rad}(\mathcal{Q}_{\mathrm{inv}})\sqrt{\kappa \log(en^2)}}{\alpha c \sqrt{2}}}$, we obtain

$$\mathbb{E}\left[\text{Term 2} \cdot \mathbb{1}\left\{\widehat{k} \geq cn\right\}\right] \leq \sqrt{\frac{\log(n)}{n}} \cdot \frac{\sqrt{\kappa}}{\alpha c \sqrt{2}} + \sqrt{\mathrm{Rad}(\mathcal{Q}_{\mathrm{inv}})\sqrt{\log(en^2)}} \cdot \frac{4\sqrt[4]{\kappa}}{\sqrt{\alpha c}}.$$

## B.4 Combining everything

First, we calculate

$$\mathrm{FDR} = \mathbb{E}\left[\mathrm{FDP}\right] = \mathbb{E}\left[\mathrm{FDP} \cdot \mathbb{1}\left\{\widehat{k} \geq c \cdot n\right\}\right] + \mathbb{E}\left[\mathrm{FDP} \cdot \mathbb{1}\left\{\widehat{k} < c \cdot n\right\}\right]$$
$$\leq \mathbb{E}\left[\mathrm{FDP} \cdot \mathbb{1}\left\{\widehat{k} \geq c \cdot n\right\}\right] + \Pr\left(\widehat{k} < c \cdot n\right),$$

since $\mathrm{FDP} \leq 1$ always. We also have $\mathrm{FDP} \leq \alpha \cdot \left[1 + \text{Term 1} + \text{Term 2}\right]$, so

$$\mathbb{E}\left[\mathrm{FDP} \cdot \mathbb{1}\left\{\widehat{k} \geq c \cdot n\right\}\right] \leq \alpha \cdot \left[1 + \mathbb{E}\left[\text{Term 1}\right] + \mathbb{E}\left[\text{Term 2} \cdot \mathbb{1}\left\{\widehat{k} \geq c \cdot n\right\}\right]\right].$$

Plugging in the bounds that we have calculated for these expected values, we have proved the theorem.

## B.5 Proof of Lemma 5

To relate the covering number to the Rademacher complexity, Srebro et al. [26, Lemmas A.2 and A.3] prove that, for $\mathcal{B} \subset [-C, C]^n$ and $\delta \leq 2C$,

$$\mathcal{N}_\infty(\mathcal{B}, \delta) \leq \left(\frac{eC\delta}{2\mathrm{Rad}(\mathcal{B})^2}\right)^{\frac{4n\mathrm{Rad}(\mathcal{B})^2}{\delta^2}}$$

as long as $\mathrm{Rad}(\mathcal{B}) < \delta/2$. Frst we see that for any $x \in \mathcal{B}$ and any index $i$,

$$\mathrm{Rad}(\mathcal{B}) \geq \frac{1}{n}\mathbb{E}\left[|\langle x, \xi\rangle|\right] \geq \frac{1}{n}\mathbb{E}\left[|\langle x, \mathbb{E}\left[\xi \mid \xi_{(-i)}\right]\rangle|\right] = \frac{1}{n}\mathbb{E}\left[|x_i \cdot \xi_i|\right] = |x_i|/n$$

by Jensen's inequality and therefore, $\mathcal{B} \subset [-n\mathrm{Rad}(\mathcal{B}), n\mathrm{Rad}(\mathcal{B})]^n$, so we can set $C = n\mathrm{Rad}(\mathcal{B})$. Assuming then that

$$2\mathrm{Rad}(\mathcal{B}) < \delta \leq 2n\mathrm{Rad}(\mathcal{B}),$$

we have

$$\mathcal{N}_\infty(\mathcal{B}, \delta) \leq \left(\frac{eC\delta}{2\mathrm{Rad}(\mathcal{B})^2}\right)^{\frac{4n\mathrm{Rad}(\mathcal{B})^2}{\delta^2}} = \left(en^2\right)^{\frac{4n\mathrm{Rad}(\mathcal{B})^2}{\delta^2}},$$

by our bounds on $C$ and $\delta$. On the other hand, if $\delta > 2C$ then we can take the covering to consist only of a single point, so the same bound holds trivially. And if $\delta \leq 2\mathrm{Rad}(\mathcal{B})$, then we can simply take a cover of the entire set $[-C, C]^n$, which contains $\mathcal{B}$:

$$\mathcal{N}_\infty(\mathcal{B}, \delta) \leq \mathcal{N}_\infty([-C, C]^n, \delta) \leq (\lceil 2C/\delta \rceil)^n \leq (\lceil 2n\mathrm{Rad}(\mathcal{B})/\delta \rceil)^n \leq (\lceil 4n\mathrm{Rad}(\mathcal{B})^2/\delta^2 \rceil)^n.$$

We also know that $a^b \leq b^a$ for any $a \geq b \geq e$, and so

$$\mathcal{N}_\infty(\mathcal{B}, \delta) \leq \left(\left\lceil \frac{4n\mathrm{Rad}(\mathcal{B})^2}{\delta^2}\right\rceil\right)^n = n^{\left\lceil \frac{4n\mathrm{Rad}(\mathcal{B})^2}{\delta^2}\right\rceil} \leq \left(en^2\right)^{\frac{4n\mathrm{Rad}(\mathcal{B})^2}{\delta^2}}.$$

Thus our bound on the covering number holds for all $\delta > 0$, which proves the lemma.

# C  Calculations for examples

In this section we prove the Rademacher complexity bounds for the three settings considered in Section 4.

## C.1  Ordered structure

Recall that for ordered (i.e. monotone) vectors $\widehat{q}$ we defined the set

$$\mathcal{Q} = \mathcal{Q}_{\text{ord}} = \{q : \epsilon \leq q_1 \leq \cdots \leq q_n \leq 1\}.$$

*Lemma 1.* For $\mathcal{Q} = \mathcal{Q}_{\text{ord}}$ as defined above,

$$\text{Rad}(\mathcal{Q}_{\text{inv}}) \leq \frac{1}{\epsilon\sqrt{n}}.$$

*Proof of Lemma 1.* For this choice of $\mathcal{Q}$, we have

$$\mathcal{Q}_{\text{inv}} = \{x : 1 \leq x_n \leq \cdots \leq x_1 \leq \epsilon^{-1}\},$$

which is the convex hull of the set

$$\mathcal{A} = \{x^0, \ldots, x^n\} \text{ where } x^k = (\underbrace{\epsilon^{-1}, \ldots, \epsilon^{-1}}_{k \text{ times}}, \underbrace{1, \ldots, 1}_{n - k \text{ times}}).$$

In this setting, since Rademacher complexity is not increased by taking a convex hull, we have (for $\xi_i \overset{\text{iid}}{\sim} \text{Unif}\{\pm 1\}$)

$$\text{Rad}(\mathcal{Q}_{\text{inv}}) = \text{Rad}(\mathcal{A}) = \frac{1}{n}\mathbb{E}\left[\max\left\{0, \max_{k=0,\ldots,n} \langle x^k, \xi \rangle\right\}\right] = \frac{1}{n}\mathbb{E}\left[\max\left\{0, \max_{k=0,\ldots,n} \epsilon^{-1}\sum_{i=1}^{k} \xi_i + \sum_{i=k+1}^{n} \xi_i\right\}\right]$$

$$= \frac{\epsilon^{-1} - 1}{n} \cdot \mathbb{E}\left[\max_{k=0,\ldots,n} \sum_{i=1}^{k} \xi_i\right] + \frac{1}{n}\mathbb{E}\left[\max\left\{0, \sum_{i=1}^{n} \xi_i\right\}\right]. \quad (27)$$

For the first term, $\mathbb{E}\left[\max_{k=0,\ldots,n} \sum_{i=1}^{k} \xi_i\right]$ is the expected maximum of a simple random walk, and it is known that

$$\mathbb{E}\left[\max_{k=0,\ldots,n} \sum_{i=1}^{k} \xi_i\right] = \sum_{k=1}^{n} k^{-1}\mathbb{E}\left[\max\left\{0, \sum_{i=1}^{k} \xi_i\right\}\right]$$

by the Pollaczek-Spitzer identity (for example this equality is implied by Borovkov [8, Section 11.8, Theorem 7]). For each $k = 1, \ldots, n$, we have

$$\mathbb{E}\left[\max\left\{0, \sum_{i=1}^{k} \xi_i\right\}\right] = \frac{1}{2}\mathbb{E}\left[\left|\sum_{i=1}^{k} \xi_i\right|\right] \leq \frac{1}{2}\sqrt{\mathbb{E}\left[\left(\sum_{i=1}^{k} \xi_i\right)^2\right]} = \frac{1}{2}\sqrt{\text{Var}\left(\sum_{i=1}^{k} \xi_i\right)} = \frac{\sqrt{k}}{2}, \quad (28)$$

where the first step holds since the distribution of $\sum_{i=1}^{k} \xi_i$ is symmetric. Thus, the first term in (27) is bounded as

$$\frac{\epsilon^{-1} - 1}{n} \cdot \mathbb{E}\left[\max_{k=0,\ldots,n} \sum_{i=1}^{k} \xi_i\right] \leq \frac{\epsilon^{-1} - 1}{n} \cdot \sum_{k=1}^{n} k^{-1} \cdot \frac{\sqrt{k}}{2}.$$

Applying (28) with $k = n$, the second term in (27) is bounded as

$$\frac{1}{n}\mathbb{E}\left[\max\left\{0, \sum_{i=1}^{n}\xi_i\right\}\right] \leq \frac{1}{n} \cdot \frac{\sqrt{n}}{2}.$$

Combining everything,

$$\text{Rad}(\mathcal{Q}_{\text{inv}}) \leq \frac{\epsilon^{-1} - 1}{2n} \cdot \sum_{k=1}^{n}\frac{1}{\sqrt{k}} + \frac{1}{2\sqrt{n}} \leq \frac{1}{\epsilon\sqrt{n}},$$

where the last step holds for any $\epsilon \leq 1$ and $n \geq 1$. □

## C.2  Group structure

Recall that we defined

$$\mathcal{Q} = \mathcal{Q}_{\text{group}} = \{q : q_i \geq \epsilon \text{ for all } i, \text{ and } q_i = q_j \text{ whenever } i, j \text{ are in the same group}\}.$$

*Lemma 2.* For $\mathcal{Q} = \mathcal{Q}_{\text{group}}$ as defined above, if the groups are of sizes $n_1, \ldots, n_d$, then

$$\text{Rad}(\mathcal{Q}_{\text{inv}}) \leq \frac{1}{2\epsilon n}\sum_{i=1}^{d}\sqrt{n_i}.$$

*Proof of Lemma 2.* In this setting, each $q \in \mathcal{Q}$ is uniquely defined by choosing the constant value inside of each group, and so taking $\xi_i \stackrel{\text{iid}}{\sim} \text{Unif}\{\pm 1\}$,

$$\text{Rad}(\mathcal{Q}_{\text{inv}}) = \frac{1}{n}\mathbb{E}\left[\sup_{x \in \mathcal{Q}_{\text{inv}}}|\langle x, \xi\rangle|\right] \leq \frac{1}{n}\mathbb{E}\left[\sup_{y \in [0,\epsilon^{-1}]^d}\sum_{i=1}^{d}\sum_{j \in \mathcal{G}_i}y_i\xi_j\right] = \frac{1}{n}\sum_{i=1}^{d}\mathbb{E}\left[\sup_{0 \leq y \leq \epsilon^{-1}} y \cdot \sum_{j \in \mathcal{G}_i}\xi_j\right]$$

$$= \frac{1}{\epsilon n}\sum_{i=1}^{d} \cdot \mathbb{E}\left[\max\left\{0, \sum_{j \in \mathcal{G}_i}\xi_j\right\}\right] \leq \frac{1}{\epsilon n}\sum_{i=1}^{d}\frac{\sqrt{n_i}}{2},$$

where the last step holds by (28). □

## C.3  Low total variation

Recall that, for the low total variation setting, we considered two choices for $\mathcal{Q}$:

$$\mathcal{Q} = \mathcal{Q}_{\text{TV-sparse}} = \left\{q \in [\epsilon, 1]^n : \sum_{(i,j) \in E}\mathbb{1}\{q_i \neq q_j\} \leq m\right\},$$

and

$$\mathcal{Q} = \mathcal{Q}_{\text{TV-}\ell_1} = \left\{q \in [\epsilon, 1]^n : \sum_{(i,j) \in E}|q_i - q_j| \leq m\right\}.$$

*Lemma 3.* For $\mathcal{Q} = \mathcal{Q}_{\text{TV-sparse}}$ as defined above,

$$\text{Rad}(\mathcal{Q}_{\text{inv}}) \leq \frac{1}{\epsilon\sqrt{n}} + \frac{2\rho_G m\sqrt{\log(n)}}{\epsilon n}.$$

If we instead take $\mathcal{Q} = \mathcal{Q}_{\text{TV-}\ell_1}$, then

$$\text{Rad}(\mathcal{Q}_{\text{inv}}) \leq \frac{1}{\epsilon\sqrt{n}} + \frac{2\rho_G m\sqrt{\log(n)}}{\epsilon^2 n}.$$

*Proof of Lemma 3.* First, for any $t_{\ell_2}, t_{\mathrm{TV}} > 0$, define

$$\mathcal{T}_G(t_{\ell_2}, t_{\mathrm{TV}}) = \left\{ x \in \mathbb{R}^n : \|x\|_2 \leq t_{\ell_2}, \sum_{(i,j) \in E_G} |x_i - x_j| \leq t_{\mathrm{TV}} \right\}.$$

Now take $\xi_i \stackrel{\mathrm{iid}}{\sim} \mathrm{Unif}\{\pm 1\}$. Hütter and Rigollet [18, Appendix B.1, equation (B.8)] calculates that

$$\max_{x \in \mathcal{T}_G(t_{\ell_2}, t_{\mathrm{TV}})} |\langle \xi, x \rangle| \leq t_{\ell_2} \cdot \|\mathcal{P}^\perp_{\mathrm{span}(D_G)}(\xi)\|_2 + t_{\mathrm{TV}} \cdot \max_{k=1,\ldots,e_G} |(D_G^+)_k^\top \xi|,$$

where $\mathrm{span}(D_G) \subseteq \mathbb{R}^n$ has co-dimension 1. Let $u$ be a unit vector orthogonal to $\mathrm{span}(D_G)$. Then

$$\mathbb{E}\left[\|\mathcal{P}^\perp_{\mathrm{span}(D_G)}(\xi)\|_2\right] = \mathbb{E}\left[|u^\top \xi|\right] \leq \sqrt{\mathbb{E}[(u^\top \xi)^2]} = 1.$$

Next, since $\xi$ is a subgaussian random vector with scale 1, $\max_{k=1,\ldots,e_G} |(D_G^+)_k^\top \xi|$ is the maximum absolute value of $e_G \leq \frac{n^2}{2}$ many centered subgaussian random variables each with scale bounded by $\rho_G = \max_k \|(D_G^+)_k\|_2$. This proves that

$$\mathrm{Rad}(\mathcal{T}_G(t_{\ell_2}, t_{\mathrm{TV}})) = \frac{1}{n} \mathbb{E}\left[\max_{x \in \mathcal{T}_G(t_{\ell_2}, t_{\mathrm{TV}})} |\langle \xi, x \rangle|\right] \leq \frac{t_{\ell_2} + t_{\mathrm{TV}} \cdot \rho_G \cdot 2\sqrt{\log(n)}}{n}. \tag{29}$$

We now analyze the complexity of $\mathcal{Q}_{\mathrm{inv}}$ in this setting. First take $\mathcal{Q} = \mathcal{Q}_{\mathrm{TV\text{-}sparse}}$ and consider any $x \in \mathcal{Q}_{\mathrm{inv}}$. Then

$$\sum_{(i,j) \in E} |x_i - x_j| = \sum_{(i,j) \in E} |(q_i)^{-1} - (q_j)^{-1}| \leq \epsilon^{-1} m,$$

since $|(q_i)^{-1} - (q_j)^{-1}| = 0$ whenever $q_i = q_j$ (i.e. for all but $m$ edges $(i,j) \in E$), and $|(q_i)^{-1} - (q_j)^{-1}| \leq \epsilon^{-1}$ for all $i, j$. We also have $\|x\|_2 \leq \epsilon^{-1}\sqrt{n}$, and so

$$\mathcal{Q}_{\mathrm{inv}} \subseteq \mathcal{T}_G(\epsilon^{-1}\sqrt{n}, \epsilon^{-1} m).$$

Applying (29), then,

$$\mathrm{Rad}(\mathcal{Q}_{\mathrm{inv}}) \leq \frac{1}{\epsilon\sqrt{n}} + \frac{2\rho_G m \sqrt{\log(n)}}{\epsilon n},$$

where $\rho_G$ is defined as before.

Next consider $\mathcal{Q} = \mathcal{Q}_{\mathrm{TV\text{-}}\ell_1}$. In this case, $\mathcal{Q}_{\mathrm{inv}}$ is actually substantially more complex: for any $x \in \mathcal{Q}_{\mathrm{inv}}$, we again have $\|x\|_2 \leq \epsilon^{-1}\sqrt{n}$, but our bound on the total variation norm is larger:

$$\sum_{(i,j) \in E} |x_i - x_j| = \sum_{(i,j) \in E} |(q_i)^{-1} - (q_j)^{-1}| = \sum_{(i,j) \in E} \frac{|q_i - q_j|}{q_i q_j} \leq \epsilon^{-2} \sum_{(i,j) \in E} |q_i - q_j| \leq \epsilon^{-2} m.$$

In this setting, then,

$$\mathcal{Q}_{\mathrm{inv}} \subseteq \mathcal{T}_G(\epsilon^{-1}\sqrt{n}, \epsilon^{-2} m).$$

Again applying (29),

$$\mathrm{Rad}(\mathcal{Q}_{\mathrm{inv}}) \leq \frac{1}{\epsilon\sqrt{n}} + \frac{2\rho_G m \sqrt{\log(n)}}{\epsilon^2 n}.$$

□

# D　Choosing $\widehat{q}$ via constrained maximum likelihood

In the different types of structure that we consider in Section 4, a common strategy for choosing $\widehat{q}$ is via the optimization problem

$$\widehat{q} = \arg\max_{q\in\mathbb{R}^n}\left\{\sum_i \mathbb{1}\{P_i > \tau\}\log(q_i(1-\tau)) + \mathbb{1}\{P_i \leq \tau\}\log(1-q_i(1-\tau)) \right.$$
$$\left. : q \in \mathcal{Q}, \sum_i \frac{\mathbb{1}\{P_i > \tau\}}{q_i(1-\tau)} \leq n \right\}. \quad (30)$$

(As before, if $\sum_i \mathbb{1}\{P_i > \tau\} > n(1-\tau)$ then we would instead set $\widehat{q} = \mathbf{1}_n$; from this point on, we proceed under the assumption that this is not the case.)

This is a constrained maximum likelihood problem, which is a convex optimization problem as long as $\mathcal{Q}$ is convex. We now give a general algorithm for this problem, implementing the Alternating Direction Method of Multipliers (ADMM) [9]. (Code for this implementation is available at http://www.stat.uchicago.edu/~rina/sabha.html.) To implement the ADMM method, we assume that the set $\mathcal{Q}$ can be characterized as follows:

$$\mathcal{Q} = \{q : Mq \in \mathcal{M}, \epsilon \leq q \leq 1\}$$

for some $\epsilon \geq 0$, some fixed matrix $M \in \mathbb{R}^{m\times n}$, and some convex set $\mathcal{M} \subseteq \mathbb{R}^m$ which has an easy-to-compute projection operator, i.e.

$$\text{Proj}_\mathcal{M}(z) = \arg\min_{x\in\mathcal{M}}\{\|x - z\|_2\}$$

is simple to compute for any $z \in \mathbb{R}^m$.

To make this concrete,

- For the ordered setting considered in Section 4.1, with $\mathcal{Q} = \mathcal{Q}_{\text{ord}}$, we take $\mathcal{M} = \{q : q_1 \leq \cdots \leq q_n\}$, and $M = \mathbf{I}_n$. The relevant projection operator can be computed via the Pool Adjacent Violators Algorithm (PAVA) [3].

- For the group-wise constant setting considered in Section 4.2, with $\mathcal{Q} = \mathcal{Q}_{\text{group}}$, we take $\mathcal{M} = \{q : q_i = q_j \text{ whenever } i, j \text{ are in the same group}\}$, and $M = \mathbf{I}_n$. The projection operator is very easy to compute: we simply take the average value within each group.

- For the bounded total variation norm setting considered in Section 4.3, suppose we want to work with $\mathcal{Q} = \mathcal{Q}_{\text{TV-}\ell_1}$, which is a convex set. Define
$$M = D_G \in \{-1, 0, +1\}^{e_G \times n},$$
the edge incidence matrix of the graph $G$ defined in Section 4.3. Then define
$$\mathcal{M} = \{z \in \mathbb{R}^{e_G} : \|z\|_1 \leq m\},$$
a rescaled $\ell_1$ unit ball; it is clear that $\mathcal{Q} = \{q : Mq \in \mathcal{M}, \epsilon \leq q \leq 1\}$, as desired. In this case, projection to $\mathcal{M}$ can be computed via soft thresholding.

## D.1　ADMM algorithm implementation

We now implement the algorithm as follows. First, our optimization problem is equivalent to calculating

$$\min_{q\in[\epsilon,1]^n, x\in\mathbb{R}^m, y\in\mathbb{R}^n}\left\{-\sum_i[\mathbb{1}\{P_i > \tau\}\log(q_i(1-\tau)) + \mathbb{1}\{P_i \leq \tau\}\log(1-q_i(1-\tau))]\right.$$
$$\left. : x \in \mathcal{M}, \sum_i \frac{\mathbb{1}\{P_i > \tau\}}{y_i(1-\tau)} \leq n, Mq = x, q = y \right\}, \quad (31)$$

which then becomes

$$\min_{q\in[\epsilon,1]^n, x\in\mathbb{R}^m, y\in\mathbb{R}^n} \max_{u\in\mathbb{R}^m, v\in\mathbb{R}^n} L(q,x,y,u,v)$$

where

$$L(q,x,y,u,v) = \left\{ -\sum_i [\mathbb{1}\{P_i > \tau\}\log(q_i(1-\tau)) + \mathbb{1}\{P_i \leq \tau\}\log(1-q_i(1-\tau))] \right.$$

$$\left. + \delta(x\in\mathcal{M}) + \delta\left(\sum_i \frac{\mathbb{1}\{P_i > \tau\}}{y_i(1-\tau)} \leq n\right) + \langle u, Mq-x\rangle + \frac{\alpha}{2}\|Mq-x\|_2^2 + \langle v, q-y\rangle + \frac{\beta}{2}\|q-y\|_2^2 \right\},$$

where $\delta(\cdot)$ is the convex indicator function and where $\alpha, \beta > 0$ are fixed parameters.
The ADMM algorithm then iterates the steps:

$$\begin{cases} q_{t+1} = \arg\min_{q\in[\epsilon,1]^n}\{L(q, x_t, y_t, u_t, v_t)\} \\ (x_{t+1}, y_{t+1}) = \arg\min_{(x,y)\in\mathbb{R}^m\times\mathbb{R}^n}\{L(q_{t+1}, x, y, u_t, v_t)\} \\ u_{t+1} = u_t + \alpha(Mq_{t+1} - x_{t+1}),\ v_{t+1} = v_t + \beta(q_{t+1} - y_{t+1}) \end{cases}$$

Now we calculate formulas for the $q$, $x$, and $y$ updates.

**The $q$ update step**  First, we modify the $q$ update step slightly: we add a preconditioning term to the $q$ update for easier computation,

$$q_{t+1} = \arg\min_{q\in[\epsilon,1]^n} \left\{ L(q, x_t, y_t, u_t, v_t) + \frac{\alpha}{2}(q-q_t)^\top(\eta\mathbf{I} - M^\top M)(q-q_t) \right\},$$

where $\eta \geq \|M\|^2$. Rearranging some terms, we are minimizing

$$-\sum_i [\mathbb{1}\{P_i > \tau\}\log(q_i(1-\tau)) + \mathbb{1}\{P_i \leq \tau\}\log(1-q_i(1-\tau))] + \frac{\alpha\eta+\beta}{2}\|q-w\|_2^2$$

over $q \in [\epsilon,1]^n$, where $w$ is the vector with entries

$$w = -\frac{M^\top(u_t + \alpha(Mq_t - x_t)) + (v_t - \beta y_t - \alpha\eta q_t)}{\alpha\eta + \beta}.$$

Note that this minimization separates over the entries $q_i$; this is the benefit of adding the preconditioning term. For $i = 1, \ldots, n$, the minimizer is given by

$$q_i = \begin{cases} \frac{w_i + \sqrt{w_i^2 + \frac{4}{\alpha\eta+\beta}}}{2} \wedge 1, & \text{if } P_i > \tau, \\ \frac{(w_i + \frac{1}{1-\tau}) - \sqrt{(w_i - \frac{1}{1-\tau})^2 + \frac{4}{\alpha\eta+\beta}}}{2} \vee \epsilon, & \text{if } P_i \leq \tau. \end{cases}$$

**The $x$ and $y$ update step**  Since the $x$ and $y$ variables do not appear jointly in any of the terms of $L(q,x,y,u,v)$, their updates are calculated independently. Trivially the $x$ update is computed as

$$x_{t+1} = \text{Proj}_{\mathcal{M}}(Mq_{t+1} + u_t/\alpha)$$

and we assume that $\mathcal{M}$ is such that this step is easy to compute. The $y$ update is given by

$$y_{t+1} = \text{Proj}_{\mathcal{G}}(q_{t+1} + v_t/\beta)$$

where $\mathcal{G} = \left\{ y \in \mathbb{R}_+^n : \sum_i \frac{\mathbb{1}\{P_i > \tau\}}{y_i(1-\tau)} \leq n \right\}$, and projection to this convex set is calculated as follows. Fix any $z \in \mathbb{R}_+^n$. If $\sum_i \frac{\mathbb{1}\{P_i > \tau\}}{z_i(1-\tau)} \leq n$ then trivially $\mathrm{Proj}_\mathcal{G}(z) = z$. If not, then for each $\lambda > 0$, define the function $f_\lambda(x)$ as the unique solution $t > x$ to the cubic equation $t^3 - t^2 x = \lambda$ (which we can calculate in closed form by the cubic formula). $\mathrm{Proj}_\mathcal{G}(z)$ is the vector $y \in \mathbb{R}_+^n$ minimizing $\frac{1}{2}\|y - z\|_2^2$ subject to $\sum_i \frac{\mathbb{1}\{P_i > \tau\}}{y_i} \leq n(1-\tau)$, so by the theory of Lagrangian multipliers, for some $\lambda > 0$ we have

$$(y - z) + \lambda \cdot (-\mathbb{1}\{P_i > \tau\}/y_i^2)_{1 \leq i \leq n} = 0.$$

In other words, for each $i$, $y_i$ satisfies $y_i - z_i - \lambda \mathbb{1}\{P_i > \tau\}/y_i^2 = 0$, i.e.

$$y_i = \begin{cases} z_i, & \text{if } P_i \leq \tau, \\ f_\lambda(z_i), & \text{if } P_i > \tau. \end{cases}$$

Now let $y(\lambda) \in \mathbb{R}_+^n$ be defined in this way, and note that $(y(\lambda))_i$ is a nondecreasing function of $\lambda$, and is strictly increasing if $P_i > \tau$. Choosing $\lambda_*$ as the unique value such that $\sum_i \mathbb{1}\{P_i > \tau\}/(y(\lambda_*))_i = n(1-\tau)$, then

$$\mathrm{Proj}_\mathcal{G}(z) = y(\lambda_*).$$